\documentclass[12pt,journal,onecolumn]{IEEEtran}
\usepackage{epsfig,rotating,setspace,latexsym,amsmath,epsf,amssymb,amsfonts,bm,theorem,subfigure,epstopdf,graphicx}
\usepackage{cite,authblk}
\usepackage{algorithmic,algorithm}
\usepackage{setspace}
\usepackage{color}
\usepackage{dsfont}
 
\newtheorem{theorem}{Theorem}
\newtheorem{lemma}{Lemma}

\newenvironment{Proof}[1]{\medskip\par\noindent{\bf Proof:\,}\,#1}{{\mbox{\,$\blacksquare$}\par}}

\allowdisplaybreaks
 
\begin{document}
 
\title{How Robust are Timely Gossip Networks to Jamming Attacks?\thanks{This paper was presented in part at the Asilomar Conference, October 2022 \cite{kaswan22jammerring} .}}
 
\author{Priyanka Kaswan \qquad Sennur Ulukus\\
        \normalsize Department of Electrical and Computer Engineering\\
        \normalsize University of Maryland, College Park, MD 20742\\
        \normalsize  \emph{pkaswan@umd.edu} \qquad \emph{ulukus@umd.edu}}
 
\maketitle

\vspace*{-2cm}

\begin{abstract}
We consider a semantics-aware communication system, where timeliness is the semantic measure, with a source which maintains the most current version of a file, and a network of $n$ user nodes with the goal to acquire the latest version of the file. The source gets updated with newer file versions as a point process, and forwards them to the user nodes, which further forward them to their neighbors using a memoryless gossip protocol. We study the average \emph{version age} of the network in the presence of $\tilde{n}$ jammers that disrupt inter-node communications, for the connectivity-constrained ring topology and the connectivity-rich fully connected topology. For the ring topology, we construct an alternate system model of mini-rings and prove that the version age of the original model can be sandwiched between constant multiples of the version age of the alternate model. We show in a ring network that when the number of jammers scales as a fractional power of the network size, i.e., $\tilde n= cn^\alpha$, the version age scales as $\sqrt{n}$ when $\alpha < \frac{1}{2}$, and as $n^{\alpha}$ when $\alpha \geq \frac{1}{2}$. As the version age of a ring network without any jammers scales as $\sqrt{n}$, our result implies that the version age with gossiping is robust against up to $\sqrt{n}$ jammers in a ring network. We then study the connectivity-rich fully connected topology, where we derive a greedy approach to place $\Tilde{n}$ jammers to maximize the age of the resultant network, which uses the jammers to isolate as many nodes as possible, thereby consolidating all links into a single mini-fully connected network. We show in this network that version age scales as $\log{n}$ when $\Tilde{n}=cn\log{n}$ and as $n^{\alpha-1}$, $1<\alpha\leq2$ when $\Tilde{n}=cn^{\alpha}$, implying that the network is robust against $n\log{n}$ jammers, since the age in a fully connected network without jammers scales as $\log{n}$. Finally, we present simulation results to support our theoretical findings.
\end{abstract}

\section{Introduction}\label{sec:intro}

We investigate the resilience of gossip-based information dissemination over networks used for timeliness purposes, against intentional jamming. We consider version age of information as a semantic metric to capture timeliness of information at the network nodes that aim to posses the latest possible version of information. In the next generation of wireless technology, goal-oriented semantic communication is going to play a critical role, wherein the meaning of the messages will be exploited in communication, timeliness being one such semantic metric. We examine the resilience of a gossip network against jamming attacks as a function of its connectivity. In particular, we consider two extremes of connectivity in symmetric networks: ring connectivity and full connectivity. In both cases, a source node which keeps the latest version of a file, updates $n$ nodes with equal update rates $\frac{\lambda}{n}$. The source itself is updated with a rate $\lambda_s$. In the case of a ring network, see Fig.~\ref{fig:ring_FC_network}(a), each node has two neighbors, and updates its neighbors with equal update rates of $\frac{\lambda}{2}$. In the case of a fully connected network, see Fig.~\ref{fig:ring_FC_network}(b), each node has $(n-1)$ neighbors, and updates its neighbors with equal update rates of $\frac{\lambda}{n-1}$. Without any jammers, in the ring network, the version age of a node scales as $\sqrt{n}$ with the network size $n$ \cite{Yates21gossip, baturalp21comm_struc}, and in the fully connected network, the version age of a node scales as $\log{n}$ \cite{Yates21gossip}. Further, in a disconnected network, where no gossiping is possible, the version age scales as $n$. We address the following questions: If there are $\tilde{n}$ jammers in a gossip network, how high can they drive the version age? What are the most and least-harmful jammer configurations? How many jammers does it take to drive the version age to order $n$, the version age with no gossiping among the nodes? We answer these questions for ring and fully connected networks.

Gossip-based algorithms are used to disseminate information efficiently in large-scale networks with no central entity to coordinate exchange of information between users. Instead, users arbitrarily contact their neighbors and exchange information based on their local status, oblivious to the simultaneous dynamics of the network as a whole, thereby causing information to spread like a gossip/rumor. Gossip algorithms are simple and scalable, and have been applied in a wide range of contexts, such as, ad-hoc routing, distributed peer sampling, autonomic self management, data aggregation, and consensus. Gossiping is introduced in \cite{Demers1987EpidemicAF-short} to fix inconsistency in clearinghouse database servers, and since then has been studied extensively, e.g., \cite{Yates21gossip_traditional,Yates21gossip,Demers1987EpidemicAF-short, Minsky02cornellthesis, vocking2000, Pittel1987OnSA, deb2006AlgebraicGossip, devavrat2006, Sanghavi2007GossipFileSplit, amazondynamo-short, Cassandra, baturalp21comm_struc, Bastopcu21gossip, kaswan22slicingcoding,kaswan22timestomping, mitra_allerton2022, elmagid23gossipagedist,Kaswan23reliable}. For example, \cite{vocking2000} shows that a single rumor can be disseminated to $n$ nodes in $O(\log n)$ rounds, \cite{deb2006AlgebraicGossip} shows that using random linear coding (RLC) $n$ messages can be disseminated to $n$ nodes in $O(n)$ time in fully connected networks, \cite{devavrat2006} further extends this result to arbitrarily connected graphs, and  \cite{Sanghavi2007GossipFileSplit} presents an improved dissemination time by dividing files into $k$ pieces.

However, data in realistic systems is not static; it keeps changing asynchronously over time as new information becomes available. For example, distributed databases like Amazon DynamoDB \cite{amazondynamo-short} and Apache Cassandra \cite{Cassandra} use gossiping for real-time peer discovery and metadata propagation. In Cassandra, cluster metadata at each node is stored in endpoint state which tracks the version number or timestamp of the data. During a single gossip exchange between two nodes, the version number of the data at the two nodes is compared, and the node with older version number discards its data in favor of the more up-to-date data present at the other node. Hence, a specific information may get discarded or lost in the network before it can reach all nodes of the network. This renders the choice of total dissemination time as a performance metric inadequate, and in such networks, the \emph{age of information} at the nodes may prove to be a more suitable performance metric. 

Age of information has been studied in a range of contexts \cite{Kosta17agesurvey, Sun19agesurvey, yates21agesurvey}. In this paper, we use \emph{version} age of information metric \cite{Yates21gossip, Eryilmaz21, bastopcu20_google} together with exponential inter-update times as used previously in \cite{bastopcu20_google, Yates17sqrt, bastopcu2020LineNetwork, kaswan_isit2021}. Version age tracks the difference between the version number of the latest file at a node and the current version prevailing at the source. Gossip networks have been studied from an age of information point of view in \cite{Yates21gossip_traditional, Yates21gossip, baturalp21comm_struc, Bastopcu21gossip, kaswan22slicingcoding, kaswan22timestomping, mitra_allerton2022, Mitra23opportunisitic,elmagid23gossipagedist,delfani22_gossip_energy, Kaswan23reliable}, where \cite{Yates21gossip_traditional,Yates21gossip} derive a recursion to find the expected age and expected version age, respectively, in arbitrary networks and characterize their scaling in fully connected graphs, \cite{baturalp21comm_struc} proves the version age for ring networks and improves the version age scaling by introducing clustering, \cite{Bastopcu21gossip} derives analogous results for the binary freshness metric,  \cite{kaswan22slicingcoding} improves version age scaling using file slicing and network coding, \cite{kaswan22timestomping} studies the effects of timestomping attacks on gossip networks, \cite{mitra_allerton2022} considers more efficient utilization of update capacity in age-aware gossiping by allowing fresher users gossip at higher rates, \cite{Mitra23opportunisitic} proposes a semi-distributed and a fully-distributed
timely gossiping scheme for fully connected networks, \cite{elmagid23gossipagedist} characterizes higher-order moments of age processes in age-aware gossip networks, \cite{delfani22_gossip_energy} considers a timely gossip network with an energy harvesting source, and \cite{Kaswan23reliable} investigates the role of reliable and unreliable sources on the age in gossiping. 

In this work, we focus on the version age in the presence of jammers  \cite{Nguyen17interferencegame, Garnaev19jamming,Xiao18jamming, Banerjee22adversary, Banerjee22game} for gossip networks. A jammer is a malicious entity that disrupts communication between two nodes, say by jamming the channel with noise. The jammer can also be a proxy for communication link failure, network partitioning, network congestion or information corruption during transfer, prevalent in distributed networks. Several works have characterized the effect of adversarial interference on gossip networks for the total dissemination time \cite{Augustine16_adversaries, Georgiou08_complexitygossip}. 

In this paper, we initiate a study of adversarial robustness of gossip networks from an age of information perspective. As an initial work in this direction, we focus on characterizing the impact of the number of jammers on the version age scaling of two types of gossip based networks, the ring network and the fully connected network, which represent the extremes of connectivity spectrum of networks. In the ring network, we show that when the number of jammers $\tilde{n}$ scales as a fractional power of network size $n$, i.e., $\tilde n= cn^\alpha$, the average version age scales with a lower bound $\Omega(\sqrt{n})$ and an upper bound $O(\sqrt{n})$ when $\alpha \in \left[0,\frac{1}{2}) \right.$, and with a lower bound $\Omega(n^{\alpha})$ and an upper bound $O(n^{\alpha})$ when $\alpha \in \left[\frac{1}{2},1\right]$, implying the version age with gossiping is robust against up to $\sqrt{n}$ jammers in a ring network, since the version age of a ring network without any jammers scales as $\sqrt{n}$. To this purpose, we construct an alternate system model of mini-rings (see Fig.~\ref{fig:best_worst_jammer_positions}) and prove that the version age of the original model can be sandwiched between constant multiples of the version age of the alternate model. Along the way, we consider average version age in line networks (see Fig.~\ref{fig:line_network_model}) and prove structural results.  

Then, we study the fully connected gossip network, where we derive a greedy approach to place $\Tilde{n}$ jammers with the goal to maximize the age of the resultant network, see Fig.~\ref{fig:engaging_1_node}. Our greedy method involves using the $\Tilde{n}$ jammers to isolate maximum possible nodes, thereby consolidating all links into a single mini-fully connected network. We show in this network that the average version age scales as $O(\log{n})$ when $\Tilde{n}=O(n\log{n})$ and as $O(n^{\alpha-1})$, $1<\alpha\leq2$ when $\Tilde{n}=O(n^{\alpha})$, implying that the network is robust against $n\log{n}$ jammers, since the version age in a fully connected network without jammers scales as $\log{n}$. 

\section{Preliminaries}

The general system model consists of a source which maintains the most up-to-date version of a file and a large network of $n$ user nodes that wish to acquire the latest version of the file. The source is updated with newer file versions with exponential inter-update times with rate $\lambda_s$. The source forwards the current file to each user node with exponential inter-update times with rate $\frac{\lambda}{n}$. Further, each user node sends updates with exponential inter-update times with rate $\lambda$ to a neighbor, chosen uniformly at random from one of its neighbors, which leads to thinning of this Poisson process into neighbor specific Poisson processes. We use $\lambda_{ij}$ to represent the rate with which node $i$ sends updates to node $j$. In addition, the system is faced with the presence of $\tilde{n}$ jammers which jam, i.e., cut, inter-node links, thereby disrupting any communication between the nodes connected by these links.

When two jammers try to cut the same link, they will together be considered as a single jammer. That is, each jammer in our model is assumed to cut a distinct inter-node link. At time $t$, if $N_i(t)$ is the latest version of a file available at user node $i$ and $N(t)$ is the current version prevailing at the source, then the instantaneous version age at node $i$ at time $t$ is $\Delta_i(t)=N(t)-N_i(t)$. The long-term expected version age of node $i$ is given by $\Delta_i=\lim_{t \to \infty} \mathbb{E}[\Delta_i(t)]$. For a set $S$ of nodes, $\Delta_S(t)=\min_{i\in S}\Delta_i(t)$, and $\Delta_S=\lim_{t \to \infty} \mathbb{E}[\Delta_S(t)]$. The expected values $\Delta_S$ are governed by \cite[Thm.~1]{Yates21gossip}, which we state here for sake of completeness, since we will need it multiple times later. Let $N(S)$ denote the set of updating neighbors of set of nodes $S$, then \cite[Thm.~1]{Yates21gossip} provides the following recursive equation,
\begin{align} \label{eqn:yates_eqn}
    \Delta_S=\frac{\lambda_s+ \sum_{i\in N(S)}(\sum_{j\in S}\lambda_{ij})\Delta_{S\cup\{i\}}}{\sum_{j\in S}\lambda_{0j}+\sum_{i\in N(S)}(\sum_{j\in S}\lambda_{ij})}
\end{align}
where $\lambda_{ij}$ is the update rate with which packets arrive from node $i$ at node $j$.

\begin{figure}[t]
 	\begin{center}
 	\subfigure[]{\includegraphics[width=0.31\linewidth]{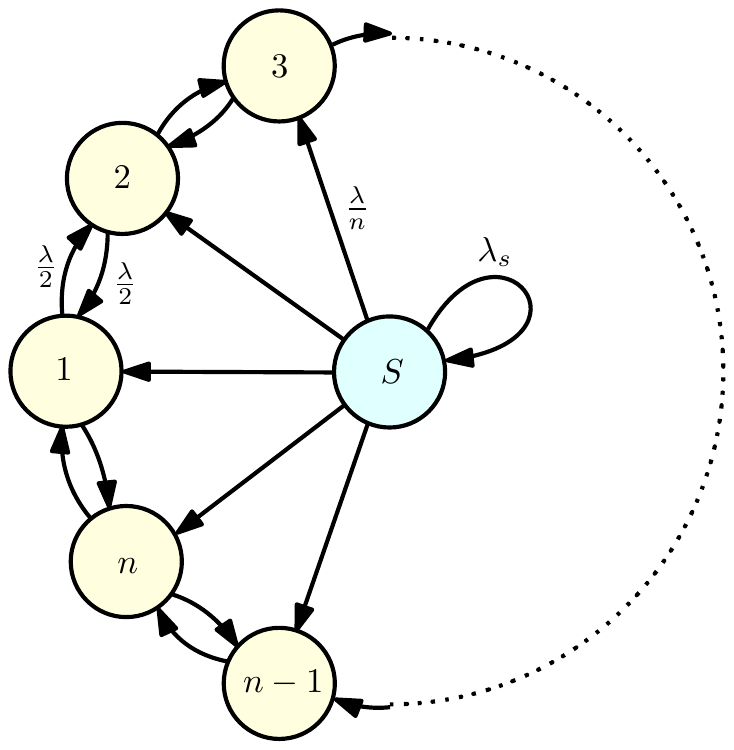}}
    \qquad \qquad
 	\subfigure[]{\includegraphics[width=0.49\linewidth]{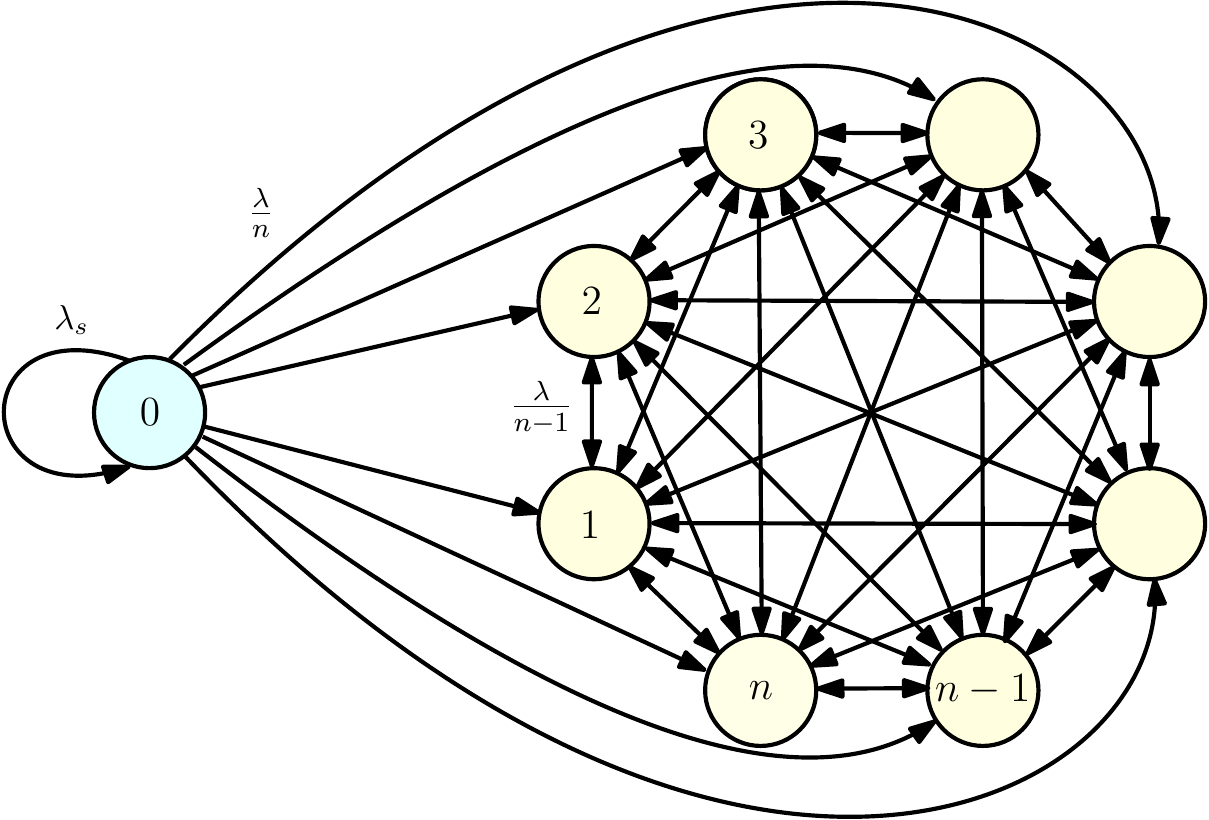}}
 	\end{center}
 	\caption{(a) Ring network of $n$ nodes. (b) Fully connected network of $n$ nodes.}
 	\label{fig:ring_FC_network}
 	\vspace{-0.4cm}
\end{figure}

In this work, we consider the two symmetric networks, the ring network in Fig.~\ref{fig:ring_FC_network}(a) and the fully connected network in Fig.~\ref{fig:ring_FC_network}(b), which represent the two extremes of network connectivity spectrum in symmetric networks. In the ring network, each user node updates each of its two nearest neighbours with exponential inter-update times with rate $\frac{\lambda}{2}$, whereas in the fully connected network, each user nodes updates every other node with exponential inter-update times with rate $\frac{\lambda}{n-1}$, such that the total update rate in both the cases is still $\lambda$ for each node. 

We begin by analyzing the ring network. When multiple adversaries cut inter-node communication links in this symmetric ring, the ring network is dismembered into a collection of isolated groups of nodes, where each group has the structure of a \emph{line network} shown in Fig.~\ref{fig:line_network_model}. The age of nodes in each such group are no longer statistically identical, owing to the disappearance of circular symmetry. In this respect, we begin by examining the spatial variation of version age over a line network of $n_0$ nodes in the next section. The analysis of a line network will be instrumental in the analysis of the ring network in subsequent sections. 

\section{Version Age in a Line Network} \label{sec:age_progress_line}

Consider the line network of $n_0$ nodes as shown in Fig.~\ref{fig:line_network_model}. In Theorem~\ref{thm:age_variation_line} below, we show that the expected version age in this network is highest at the corner nodes and decreases towards the center. Intuitively, this is due to the fact that corner nodes are updated less frequently as they are connected with only one inter-node link, and thus, the exchanges towards the network corners are based on relatively staler file versions. Here, superscript $\ell(n_0)$ denotes a line network of size $n_0$.

\begin{figure}[t]
\centerline{\includegraphics[width=0.53\linewidth]{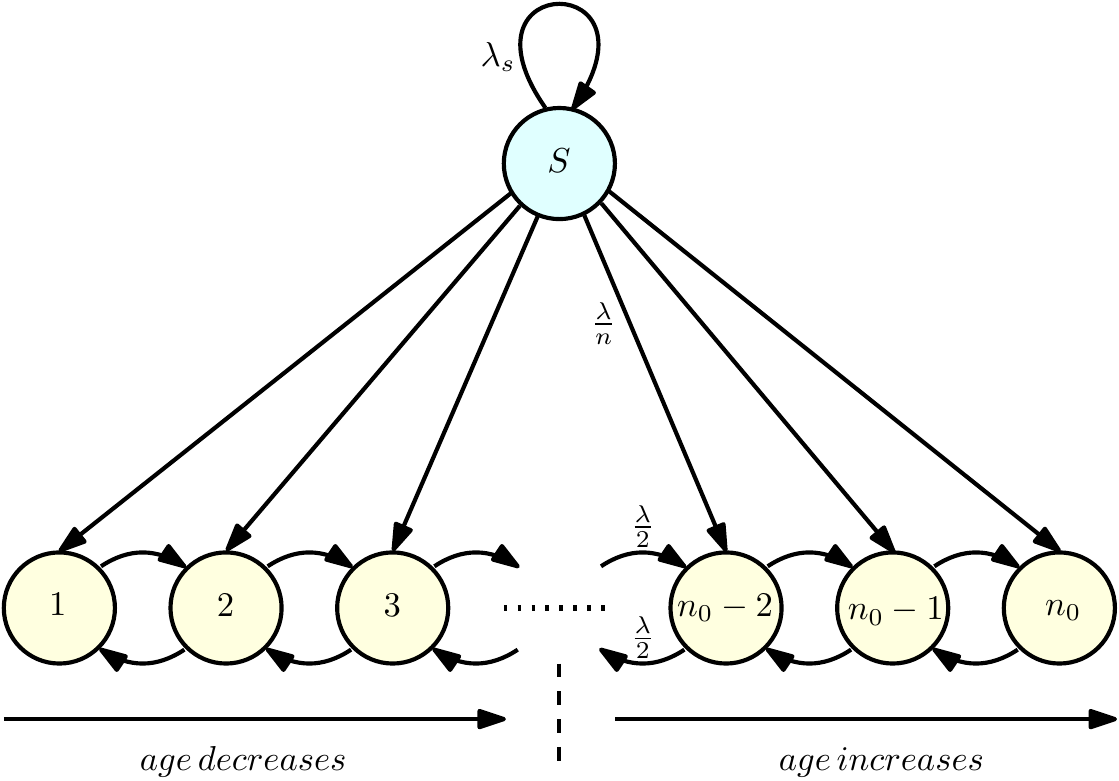}}
\caption{Line network model with $n_0$ nodes.}
\label{fig:line_network_model}
\vspace*{-0.4cm}
\end{figure}

\begin{theorem}\label{thm:age_variation_line}
    In a line network, $\Delta^{\ell(n_0)}_{i+1} \leq \Delta^{\ell(n_0)}_i$,  $i\leq\frac{n_0}{2}$. 
\end{theorem}

\begin{figure}[t]
\centerline{\includegraphics[width=0.7\linewidth]{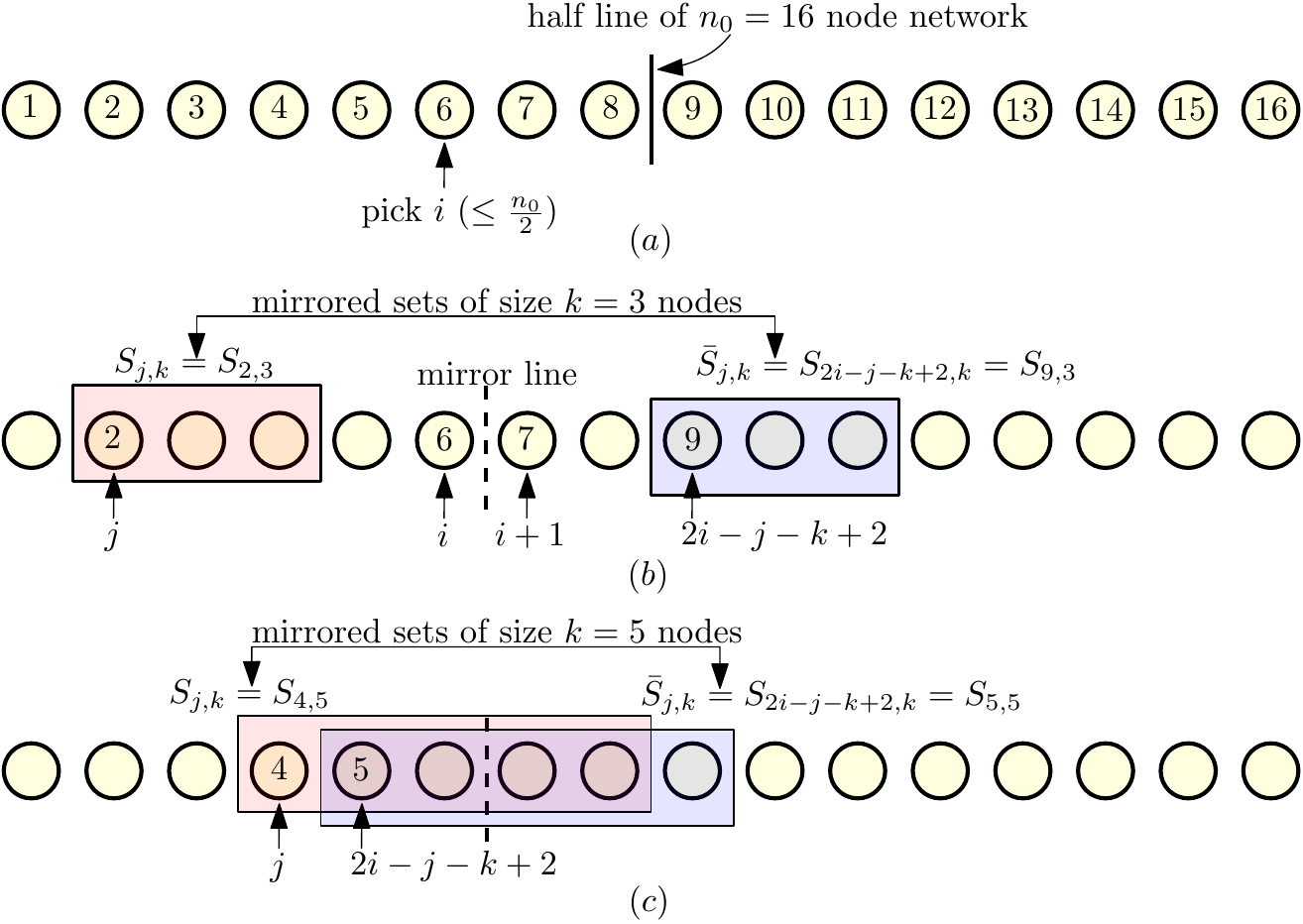}}
\caption{$S_{j,k}$ (pink blocks) and $\bar{S}_{j,k}=S_{2i-j-k+2,k}$ (blue blocks) positioned symmetrically about the dotted line between nodes $i$ and $i+1$.}
\label{fig:line_network_blocks}
\vspace*{-0.4cm}
\end{figure}

\begin{Proof}
Fix the $i$ in the statement of the theorem. Let $S_{j,k}=\{j,\ldots,j+k-1\}$ denote a set of $k$ contiguous nodes, beginning with node $j$, in a size $n_0$ line network, where $j+k-1\leq n_0$, see Fig.~\ref{fig:line_network_blocks}. We use $\Delta^{\ell(n_0)}_{j,k}$ to denote $\Delta^{\ell(n_0)}_{S_{j,k}}$, i.e., replace the set with its indices. Define $\bar{S}_{j,k}$ as the mirror image of set $S_{j,k}$ about the dotted line between nodes $i$ and $i+1$. Note that $\bar{S}_{j,k}=S_{2i-j-k+2,k}$; see Fig.~\ref{fig:line_network_blocks} for examples. Similarly, we use $\bar{\Delta}^{\ell(n_0)}_{j,k}$ to denote $\Delta^{\ell(n_0)}_{\bar{S}_{j,k}}$. We will consider sets $S_{j,k}$ where majority of the elements of $S_{j,k}$ lie to the left of node $i$, i.e., $j\leq i+1-\frac{k}{2}$, and prove that $\bar{\Delta}^{\ell(n_0)}_{j,k}\leq \Delta^{\ell(n_0)}_{j,k}$. Then, taking $k=1$ (size one set) with $j=i$ gives the desired result.  

We provide a proof by induction, beginning with the case $k=2i$, where $j=1$ is the only value that meets the condition $j\leq i+1-\frac{k}{2}$. This case gives $\bar{S}_{j,k} = S_{j,k}$ and $\bar{\Delta}^{\ell(n_0)}_{j,k}=\Delta^{\ell(n_0)}_{j,k}$ which satisfies the claim. Next, we assume that the claim holds for some $k$, i.e., $\bar{\Delta}^{\ell(n_0)}_{j,k}\leq \Delta^{\ell(n_0)}_{j,k}$ for all $j\leq i+1-\frac{k}{2}$, and prove it for $k-1$. To relate version ages of size $k$ and $k-1$ sets, we apply (\ref{eqn:yates_eqn}) to obtain the version age of $S_{j,k-1}$
\begin{align}
\Delta^{\ell(n_0)}_{j,k-1}= \begin{cases} \label{eqn:k-1_block_age_trans_cases}
\frac{\lambda_s+\frac{\lambda}{2}\Delta^{\ell(n_0)}_{j,k}+\frac{\lambda}{2}\Delta^{\ell(n_0)}_{j-1,k}}{\frac{(k-1)\lambda}{n}+\lambda}, & j> 1\\
\frac{\lambda_s+\frac{\lambda}{2}\Delta^{\ell(n_0)}_{j,k}}{\frac{(k-1)\lambda}{n}+\frac{\lambda}{2}},& j=1
\end{cases}
\end{align}
and for its mirrored set $\bar{S}_{j,k-1}$ as
\begin{align} \label{eqn:k-1_mirrorblock_age_trans}
    \bar{\Delta}^{\ell(n_0)}_{j,k-1}=\frac{\lambda_s+\frac{\lambda}{2} \bar{\Delta}^{\ell(n_0)}_{j,k} +\frac{\lambda}{2} \bar{\Delta}^{\ell(n_0)}_{j-1,k}}{\frac{(k-1)\lambda}{n}+\lambda}
\end{align}
Since $\bar{S}_{j,k-1} \subset \bar{S}_{j-1,k}$, we have $\bar{\Delta}^{\ell(n_0)}_{j,k-1} \geq  \bar{\Delta}^{\ell(n_0)}_{j-1,k}$ as an extended feasible region provides a lower minimum. Using this inequality to eliminate the last term in (\ref{eqn:k-1_mirrorblock_age_trans}) gives
\begin{align} \label{eqn:k-1_mirrorblock_age_trans_inequality}
    \bar{\Delta}^{\ell(n_0)}_{j,k-1} \leq \frac{\lambda_s+\frac{\lambda}{2}\bar{\Delta}^{\ell(n_0)}_{j,k}}{\frac{(k-1)\lambda}{n}+\frac{\lambda}{2}}
\end{align}

We wish to prove the claim for $k-1$, i.e., $\bar{\Delta}^{\ell(n_0)}_{j,k-1} \leq \Delta^{\ell(n_0)}_{j,k-1}$ under the condition $j\leq i+1-\frac{(k-1)}{2}$. If $j=i+1-\frac{(k-1)}{2}$, then we get $j=2i-j-k+3$, giving $\bar{S}_{j,k-1} = S_{j,k-1}$ and $\bar{\Delta}^{\ell(n_0)}_{j,k-1} = \Delta^{\ell(n_0)}_{j,k-1}$ which satisfies the claim. For other $j$, as $j$ must be integer valued, $j\leq i+1-\frac{(k-1)}{2}$ is automatically implied from  $j\leq i+1-\frac{k}{2}$. Consequently, since the claim is assumed to hold for $k$, we have $\bar{\Delta}^{\ell(n_0)}_{j,k}\leq \Delta^{\ell(n_0)}_{j,k}$ and $\bar{\Delta}^{\ell(n_0)}_{j-1,k}\leq \Delta^{\ell(n_0)}_{j-1,k}$. Now, comparing the terms in (\ref{eqn:k-1_block_age_trans_cases}) and (\ref{eqn:k-1_mirrorblock_age_trans}) for case $j>1$, and (\ref{eqn:k-1_block_age_trans_cases}) and (\ref{eqn:k-1_mirrorblock_age_trans_inequality}) for case $j=1$ gives $\bar{\Delta}^{\ell(n_0)}_{j,k-1} \leq \Delta^{\ell(n_0)}_{j,k-1}$, and thus, the claim holds for $k-1$. Finally, setting $k=1$ and $j=i$, we have $\Delta^{\ell(n_0)}_{i+1}=\Delta^{\ell(n_0)}_{i+1,1}=\bar{\Delta}^{\ell(n_0)}_{i,1}\leq \Delta^{\ell(n_0)}_{i,1}=\Delta^{\ell(n_0)}_i$, which completes the proof.
\end{Proof}

\section{Bounds on Age of Line Network} \label{sec:bounds_line}

\begin{figure}[t]
\centerline{\includegraphics[width=0.6\linewidth]{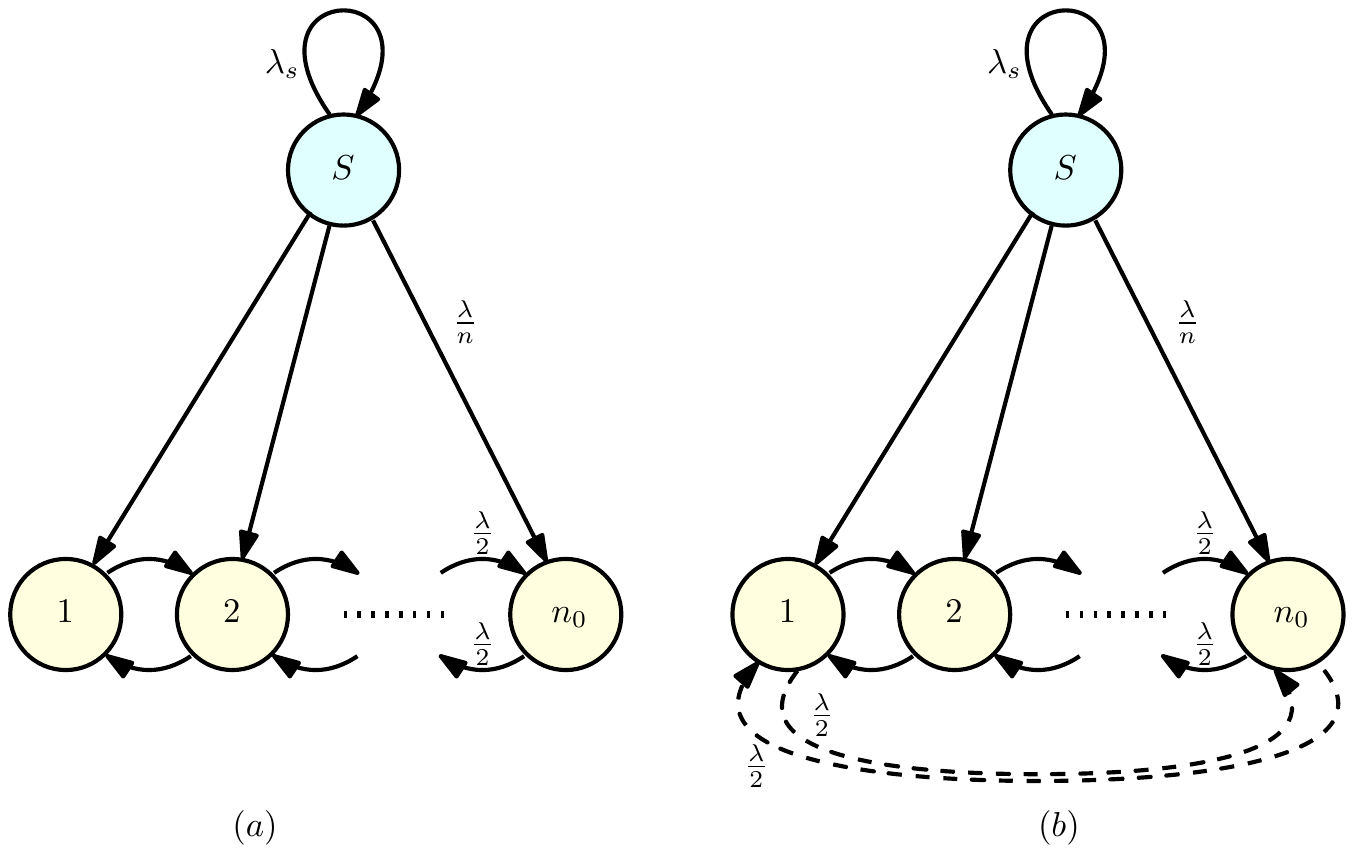}}
\caption{(a) A line network of size $n_0$. (b) A ring network of size $n_0$. Compared to (a), an extra link is introduced between end nodes bringing radial symmetry.}
\label{fig:line_to_miniring}
\vspace*{-0.4cm}
\end{figure}

Here, we bound $\Delta^{\ell(n_0)}_{i}$, age in a line network of size $n_0$ with $\Delta^{r(n_0)}_{i}$, age in a ring network of size $n_0$, see Fig.~\ref{fig:line_to_miniring}. 

\subsection{Lower Bound} \label{subsec:lowerbound_line}

Heuristically, due to the presence of an additional link compared to the line network, the ring network will have more age-conformed transitions, and therefore, improved age at the nodes. Mathematically, the recursive equation in (\ref{eqn:yates_eqn}) is identical in the two networks for every subset of nodes $S_{j,k} \in \{2,\ldots,n_0-1\}$ that excludes corner nodes $1$ and $n_0$. 

Further, for subsets $S_{1,k}$ which include corner node $1$, similar to  (\ref{eqn:k-1_block_age_trans_cases}) for case $j=1$, we have
\begin{align} \label{eqn:line_end_node}
    \Delta^{\ell(n_0)}_{1,k}=\frac{\lambda_s+\frac{\lambda}{2}\Delta^{\ell(n_0)}_{1,k+1}}{\frac{k\lambda}{n}+\frac{\lambda}{2}}
\end{align}
The corresponding equation for the ring, similar to (\ref{eqn:k-1_mirrorblock_age_trans})-(\ref{eqn:k-1_mirrorblock_age_trans_inequality}), is
\begin{align}
    \!\!\! \Delta^{r(n_0)}_{1,k}=\frac{\lambda_s+\frac{\lambda}{2}\Delta^{r(n_0)}_{1,k+1}+\frac{\lambda}{2}\Delta^{r(n_0)}_{n_0,k+1}}{\frac{k\lambda}{n}+\lambda}
    \leq \frac{\lambda_s+\frac{\lambda}{2}\Delta^{r(n_0)}_{1,k+1}}{\frac{k\lambda}{n}+\frac{\lambda}{2}} \label{eqn:ring_end_node}
\end{align}
where $S_{n_0,k+1}$ refers to the set $\{n_0,1,2,\ldots,k\}$, and hence, $S_{1,k}\subset S_{n_0,k+1}$ gives the inequality $\Delta^{r(n_0)}_{n_0,k+1} \leq \Delta^{r(n_0)}_{1,k}$.

Now, comparing (\ref{eqn:line_end_node}) and (\ref{eqn:ring_end_node}) gives $\Delta^{r(n_0)}_{1,k} \leq \Delta^{\ell(n_0)}_{1,k}$. By symmetry, $ \Delta^{r(n_0)}_{n_0+1-k,k} \leq \Delta^{\ell(n_0)}_{n_0+1-k,k}$ for subsets which include the other corner node $n_0$. Thus, $\Delta^{r(n_0)}_{j,k} \leq \Delta^{\ell(n_0)}_{j,k}$, for all $j,k$, and $\Delta^{r(n_0)}_{i} \leq \Delta^{\ell(n_0)}_{i}$, which gives us a lower bound on $\Delta^{\ell(n_0)}_{i}$.

\subsection{Upper Bound}

Note that due to the radial symmetry of the ring network, $\Delta^{r(n_0)}_i=\Delta^{r(n_0)}_1$, for all $i$. Hence, combining results of Sections~\ref{sec:age_progress_line}~and~\ref{subsec:lowerbound_line}, with bilateral symmetry of line network, 
\begin{align}
    \Delta^{r(n_0)}_{1} &\leq \Delta^{\ell(n_0)}_{\lceil\frac{n_0}{2}\rceil}=\Delta^{\ell(n_0)}_{\lfloor\frac{n_0}{2}\rfloor}\leq \ldots \leq\Delta^{\ell(n_0)}_{n_0-1}=\Delta^{\ell(n_0)}_{2}\leq \Delta^{\ell(n_0)}_{n_0}=\Delta^{\ell(n_0)}_{1} 
\end{align}
Hence, an upper bound on $\Delta^{\ell(n_0)}_{1}$ is an upper bound on $\Delta^{\ell(n_0)}_{i}$ for all $i$. Recursively applying (\ref{eqn:yates_eqn}), see also \cite[Lem.~2]{baturalp21comm_struc},
\begin{align}\label{eqn:ring_end_node_detailage}
    \Delta^{r(n_0)}_1=\frac{\lambda_s}{\lambda}\Bigg[\sum_{j=1}^{n_0-1}\prod_{k=1}^{j}\frac{1}{\frac{k}{n}+1}+\frac{1}{\frac{n_0}{n}}\prod_{k=1}^{n_0-1}\frac{1}{\frac{k}{n}+1}\Bigg]
\end{align}
and
\begin{align}
    \Delta^{\ell(n_0)}_1=&\frac{2\lambda_s}{\lambda} \Bigg[\sum_{j=1}^{n_0-1}\prod_{k=1}^{j}\frac{1}{\frac{k}{n/2}+1} +\frac{1}{\frac{n_0}{n/2}}\prod_{k=1}^{n_0-1}\frac{1}{\frac{k}{n/2}+1}\Bigg]\\
    \leq & \frac{\lambda_s}{\lambda}\Bigg[\sum_{j=1}^{n_0-1}\prod_{k=1}^{j}\frac{1}{\frac{k}{n}+1}+\frac{1}{\frac{n_0}{n}}\prod_{k=1}^{n_0-1}\frac{1}{\frac{k}{n}+1}\Bigg] \\
    =&2\Delta^{r(n_0)}_1 
\end{align}

Hence for each node $i$, 
\begin{align}
\Delta^{r(n_0)}_1\leq \Delta^{\ell(n_0)}_i \leq 2\Delta^{r(n_0)}_1
\end{align}
Since $\Delta^{\ell(n_0)}_i$ is sandwiched between constant multiples of $\Delta^{r(n_0)}_1$, they both scale similarly when $n$ is large. Intuitively, when the number of nodes is large, the effect of one additional link in the \emph{line} versus \emph{ring} models becomes insignificant.

\section{Jammer Positioning for Age Deterioration in Ring Network} \label{subsec:position_jammers}

To characterize the total age of the system, we define $\Delta^{\ell(n_0)}=\sum_{i=1}^{n_0}\Delta^{\ell(n_0)}_i$ and $\Delta^{r(n_0)}=\sum_{i=1}^{n_0}\Delta^{r(n_0)}_i$, the sum of version ages in size $n_0$ line and ring networks, respectively. 

\begin{lemma}\label{lemma:ring_age_difference}
$\Delta^{r(n_0)} -\Delta^{r(n_0+1)}$ decreases with increase in $n_0$.
\end{lemma}

\begin{Proof}
We have $\Delta^{r(n_0)}=n_0 \Delta^{r(n_0)}_1$ due to the radial symmetry of a ring. Hence, by obtaining expressions for $\Delta^{r(n_0)}_1$ and $\Delta^{r(n_0+1)}_1$ from (\ref{eqn:ring_end_node_detailage}), we get
\begin{align}
    \Delta^{r(n_0)}-\Delta^{r(n_0+1)}=& n_0\Delta^{r(n_0)}_1-(n_0+1)\Delta^{r(n_0+1)}_1 \\
    =&\frac{\lambda_s}{\lambda}\Bigg[-\sum_{j=1}^{n_0}\prod_{k=1}^{j}\frac{1}{\frac{k}{n}+1}\Bigg] 
\end{align}
which decreases with increasing $n_0$.
\end{Proof}

\begin{figure}[t]
\centerline{\includegraphics[width=0.85\linewidth]{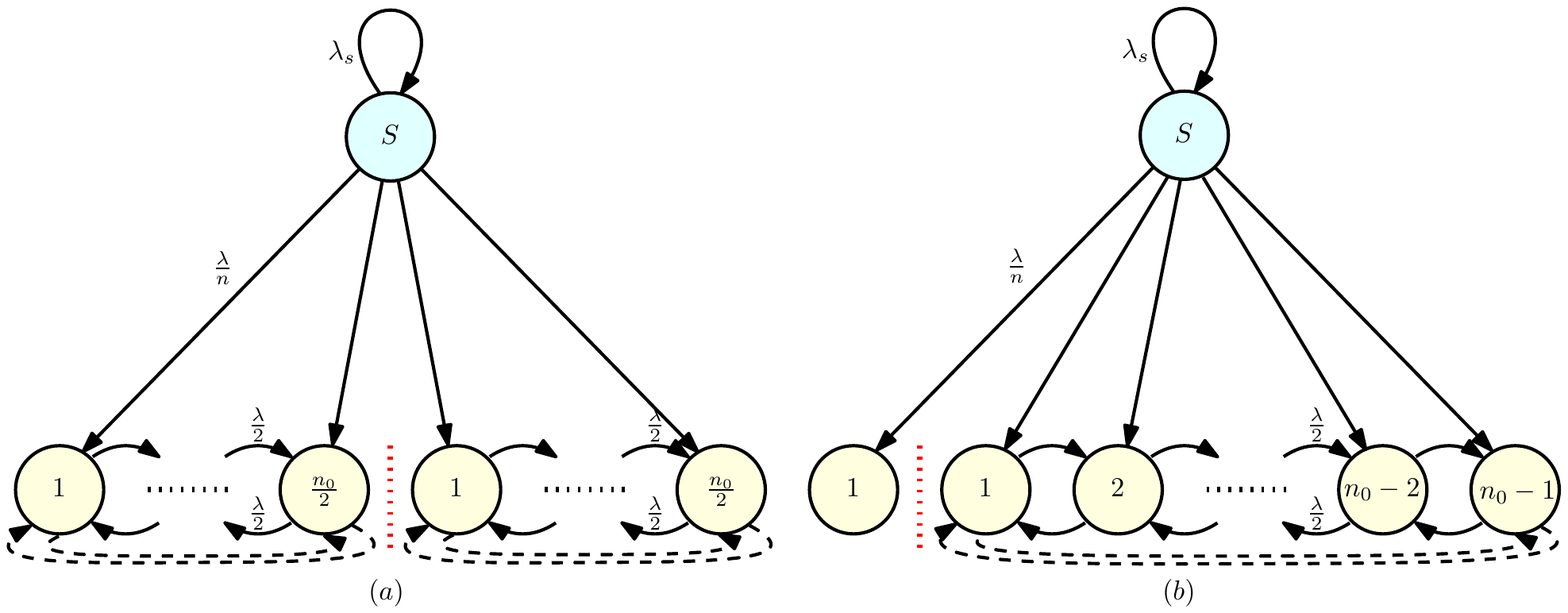}}
\caption{Jammer position on line network (a) most favorable, (b) most harmful.}
\label{fig:best_worst_rings}
\vspace*{-0.4cm}
\end{figure}

Next, we consider the problem where a single jammer cuts a single link in the line network model of Fig.~\ref{fig:line_network_model}. Let the jammer cut the link between nodes $m$ and $m+1$, effectively converting the line network of size $n_0$ into two smaller line networks of sizes $m$ and $n_0-m$. The age of the resulting system is $\Delta^{\ell(n_0)(m)}=\Delta^{\ell(m)} +\Delta^{\ell(n_0-m)}$. We will consider the mini-ring approximation to the resulting two line networks, and assume that the end points of the individual line networks are connected to form mini-rings as shown in Fig.~\ref{fig:best_worst_rings}. The age of this network with mini-rings is $\Delta^{r(n_0)(m)}=\Delta^{r(m)} +\Delta^{r(n_0-m)}$. From Section~\ref{sec:bounds_line}, the total ages of the actual dismembered line network and the mini-ring approximation are related as $\Delta^{r(n_0)(m)} \leq \Delta^{\ell(n_0)(m)}\leq 2\Delta^{r(n_0)(m)}$. Next, as an approximation to finding $m$ that maximizes (worst case jammer) and minimizes (most favorable jammer) $\Delta^{\ell(n_0)(m)}$, we will find $m$ that maximizes/minimizes $\Delta^{r(n_0)(m)}$ instead. We show in Theorem~\ref{thm:best_position_for_jammer} below that most harmful jammer cuts the link that separates the node at the corner, and the most favorable jammer cuts the center link, as shown in Fig.~\ref{fig:best_worst_rings}.

\begin{theorem} \label{thm:best_position_for_jammer}
In a ring, $\Delta^{r(n_0)(m+1)}\leq \Delta^{r(n_0)(m)}$, $m \leq \frac{n_0}{2}$.
\end{theorem}

\begin{Proof}
We have 
\begin{align}
    \Delta^{r(n_0)(m)}-\Delta^{r(n_0)(m+1)} &= \left[\Delta^{r(m)} +\Delta^{r(n_0-m)}\right]- \left[\Delta^{r(m+1)} +\Delta^{r(n_0-m-1)}\right] \\
    &= \left[\Delta^{r(m)} -\Delta^{r(m+1)}\right] - \left[\Delta^{r(n_0-m-1)} -\Delta^{r(n_0-m)}\right]\\
    &\geq 0 
\end{align}
where the last inequality follows from Lemma~\ref{lemma:ring_age_difference}, as $m<n_0-m-1$ since we are only considering $m\leq \frac{n_0}{2}$.
\end{Proof}

\begin{figure}[t]
\centerline{\includegraphics[width=0.7\linewidth]{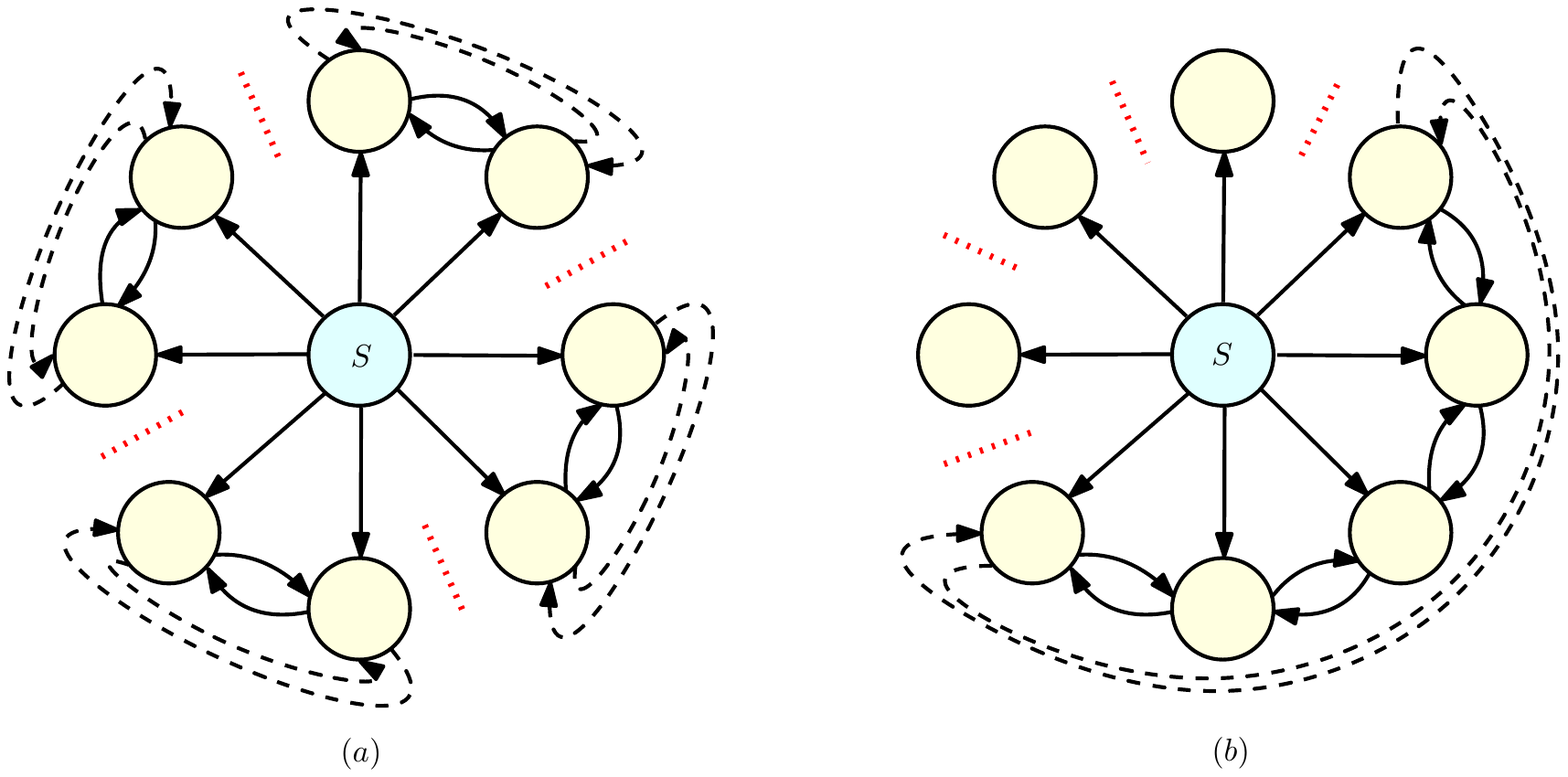}}
\caption{Jammer positions on a ring (a) most favorable, (b) most harmful.}
\label{fig:best_worst_jammer_positions}
\vspace*{-0.4cm}
\end{figure}

Consider the original ring network of $n$ nodes in Fig.~\ref{fig:ring_FC_network}(a) in the presence of $\tilde{n}$ jammers cutting $\tilde{n}$ communication links. This results in $\tilde{n}$ isolated line networks. Let $\Delta^\ell$ denote the average version age at in this dismembered ring, which is composed of $\tilde{n}$ line networks. Consider the alternate model, where all the $\tilde{n}$ line networks are replaced by their mini-ring versions, as shown in Fig.~\ref{fig:best_worst_jammer_positions}(a), and let $\Delta^r$ denote the average age of this alternate system. Then, from Section~\ref{sec:bounds_line}, we know that $\Delta^\ell$ is bounded by constant multiples of $\Delta^r$.

From Theorem~\ref{thm:best_position_for_jammer}, the least harmful positioning of jammers for $\Delta^r$ is the equidistant placement around the ring, as shown in Fig.~\ref{fig:best_worst_jammer_positions}(a), since we can keep switching a jammer's position until it has equal size mini-rings on both sides. Likewise, the most detrimental positioning of jammers is if they cut adjacent links, as shown in Fig.~\ref{fig:best_worst_jammer_positions}(b), which results in $\tilde{n}-1$ isolated nodes and a single line network of $n-\tilde{n}+1$ nodes. This is because from Theorem~\ref{thm:best_position_for_jammer}, presence of two line networks of $i_1$ and $i_2$ nodes has lower age than presence of a single line network of $i_1+i_2-1$ nodes and an isolated node.

\section{Average Age of Ring Network with Jammers} \label{sec:avg_age_ring}

\begin{lemma}\label{lemma:gaussian_approx} (a)
$\sum_{j=1}^{n_0}\left[\prod_{k=1}^{j}\frac{1}{\frac{k}{n}+1}\right]=O(\sqrt{n})$. (b) If $n_0=\omega(\sqrt{n})$, then $\sum_{j=1}^{n_0}\left[\prod_{k=1}^{j}\frac{1}{\frac{k}{n}+1}\right]=\Omega(\sqrt{n})$.
\end{lemma}
\begin{Proof}
We use $\frac{x}{1+x} \leq \log(1+x) \leq x$ and $\sum_{k=1}^{j} \frac{k}{n} = \frac{j(j+1)}{2n}$ to bound $\sum_{k=1}^{j} \log \big( 1+\frac{k}{n} \big)$ and then exponentiate to obtain
\begin{align}\label{eqn:negative_exponent_bounds}
    e^{-\frac{j^2}{n}} \leq  \prod_{k=1}^{j} \frac{1}{1+\frac{k}{n}}  \leq e^{-\frac{j^2}{4n}}
\end{align}
Then, we sum over $j$ and use Riemann sums to obtain
\begin{align}
    \int_{\frac{1}{\sqrt{n}}}^{\frac{n_0}{\sqrt{n}}}e^{-\frac{t^2}{C}}dt \leq \frac{1}{\sqrt{n}}\sum_{j=1}^{n_0}  e^{-\frac{j^2}{Cn}} \leq \int_{0}^{\frac{n_0}{\sqrt{n}}}e^{-\frac{t^2}{C}}dt 
\end{align}
and note that $\int_{0}^{\infty}e^{-\frac{t^2}{C}}dt = \frac{\sqrt{C\pi}}{2}$ for a constant $C>0$.
\end{Proof}

\subsection{Lower Bound on System Age}

\begin{theorem} \label{thm:lowerbound}
For a ring network of $n$ nodes with $\tilde{n} =c n^{\alpha}$ jammers, where $\alpha \in [0,1]$ and $c$ is a scaling constant,
\begin{align}
\Delta^{\ell} = \begin{cases} 
\Omega(n^{\alpha}), & \alpha\geq \frac{1}{2}\\
\Omega(\sqrt{n}), & \alpha<\frac{1}{2}
\end{cases}
\end{align}
\end{theorem}

\begin{Proof}
Based on Section~\ref{subsec:position_jammers}, to lower bound $\Delta^r(\leq \Delta^{\ell})$, we consider the least detrimental model of Fig.~\ref{fig:best_worst_jammer_positions}(a), where each node is a part of $n_0=\frac{n}{\tilde n}=\frac{1}{c}n^{1-\alpha}$ size mini-ring, and hence all nodes have identical average age giving $\Delta^r=\Delta^{r(n_0)}_1$.

For $\alpha<\frac{1}{2}$, using Lemma~\ref{lemma:gaussian_approx}(b) in (\ref{eqn:ring_end_node_detailage}) gives
\begin{align} \label{eqn:lowerbound_alpha3}
    \!\!\!\Delta^{r(n_0)}_1 \!\geq\! \frac{\lambda_s}{\lambda}\Bigg[\sum_{j=1}^{n_0-1}\!\prod_{k=1}^{j}\frac{1}{\frac{k}{n}+1}\!+\!\frac{n}{n_0}\!\prod_{k=1}^{n_0}\frac{1}{\frac{k}{n}+1}\Bigg] 
    \!\geq\!\Omega(\sqrt{n}) 
\end{align}
For $\alpha \geq \frac{1}{2}$, using (\ref{eqn:negative_exponent_bounds}) in (\ref{eqn:ring_end_node_detailage}) gives
\begin{align} \label{eqn:lowerbound_alpha8}
    \Delta^{r(n_0)}_1 \geq &\frac{\lambda_s}{\lambda}\left[0 +c n^{\alpha}e^{-\frac{n_0^2}{n}} \right] =\Omega(n^{\alpha})
\end{align}
where $e^{-\frac{n_0^2}{n}}= e^{-\frac{n^{-(2\alpha-1)}}{2c^2}}\to 1$ for large $n$.
\end{Proof}

\subsection{Upper Bound on System Age}
\begin{theorem} \label{thm:upperbound}
For a ring network of $n$ nodes with $\tilde{n} =c n^{\alpha}$ jammers, where $\alpha \in (0,1)$ and $c$ is a scaling constant, 
\begin{align}
\Delta^{\ell} = \begin{cases} 
O(n^{\alpha}), & \alpha\geq \frac{1}{2}\\
O(\sqrt{n}), & \alpha<\frac{1}{2}
\end{cases} 
\end{align}
\end{theorem}

\begin{Proof}
To upper bound $\Delta^r(\geq \frac{1}{2}\Delta^{\ell})$, we consider the most detrimental model of Fig.~\ref{fig:best_worst_jammer_positions}(b), which has $\tilde n -1$ isolated nodes and a line network of $n-\tilde{n} +1$ nodes. Then,
\begin{align} \label{eqn:average_worstcase_jammer}
    \Delta^r&= \frac{(\tilde n -1)\Delta^{r(1)}_1 + (n-\tilde n +1) \Delta^{r(n-\tilde{n} +1)}_1}{n} 
\end{align}
Since $\frac{\tilde n-1}{n} \leq \frac{\tilde n}{n}$ and $\frac{n-\tilde n+1}{n}=\frac{n(1-cn^{-(1-\alpha)}+\frac{1}{n})}{n} \leq 1$, using (\ref{eqn:ring_end_node_detailage}), (\ref{eqn:negative_exponent_bounds}) and Lemma~\ref{lemma:gaussian_approx}(a) in (\ref{eqn:average_worstcase_jammer}) gives
\begin{align} \label{eqn:upperbound}
    \Delta^r \leq& \frac{\tilde n}{n}\frac{\lambda_s n}{\lambda}+\frac{n-\tilde n+1}{n} \frac{\lambda_s}{\lambda} \left[ O(\sqrt{n}) + \frac{n c e^{-\frac{(n-\tilde n)^2}{4n}} }{n-\tilde n+1}\right] \\
    \leq&\frac{\lambda_s}{\lambda}\left[cn^{\alpha} + O(\sqrt{n}) + c \right] 
\end{align}
completing the proof.
\end{Proof}

\section{Fully Connected Network}\label{sec:FC}

In the previous sections, we focused on the ring network, where each node only communicates with its neighbors on both sides with rates $\frac{\lambda}{2}$, and hence, there are fewer links in the network carrying a larger load of information. In contrast, the fully connected network exhibits the other extreme of network connectivity, where every node communicates with every other node of the network where all links transmit information at a lower rate of $\frac{\lambda}{n-1}$; Fig.~\ref{fig:ring_FC_network}(b). We saw that version age scaling in a ring network is robust up to $O(\sqrt{n})$ jammers. One would expect jammers to have more success in the ring network compared to the fully connected network, since jamming any link blocks a higher number of transitions in a ring. To investigate this intuition, the first step is to find what the worst case configuration for $\Tilde{n}$ jammers in a fully connected network of $n$ nodes is, and if the robustness of age scaling against jamming is stronger than in ring networks. 

To this end, we consider an alternate fully connected network, where inter-node communications happen at rate $\frac{\lambda}{n}$ instead of $\frac{\lambda}{n-1}$. Naturally, the age of this modified network, which we will henceforth simply call the fully connected network, is going to be an upper bound to the age of the \emph{original} fully connected network, since the links now have lower update rates, translating to fewer age-minimizing updates on all the links, which can be verified from (\ref{eqn:yates_eqn}). Therefore, the robustness of age scaling of this upper bound would imply robustness of age scaling in the original network, and in the regime of large $n$, with $\frac{\lambda}{n-1} \approx \frac{\lambda}{n}$, expected ages in both the networks would converge to each other.

\begin{figure}[t]
\centerline{\includegraphics[width=0.95\linewidth]{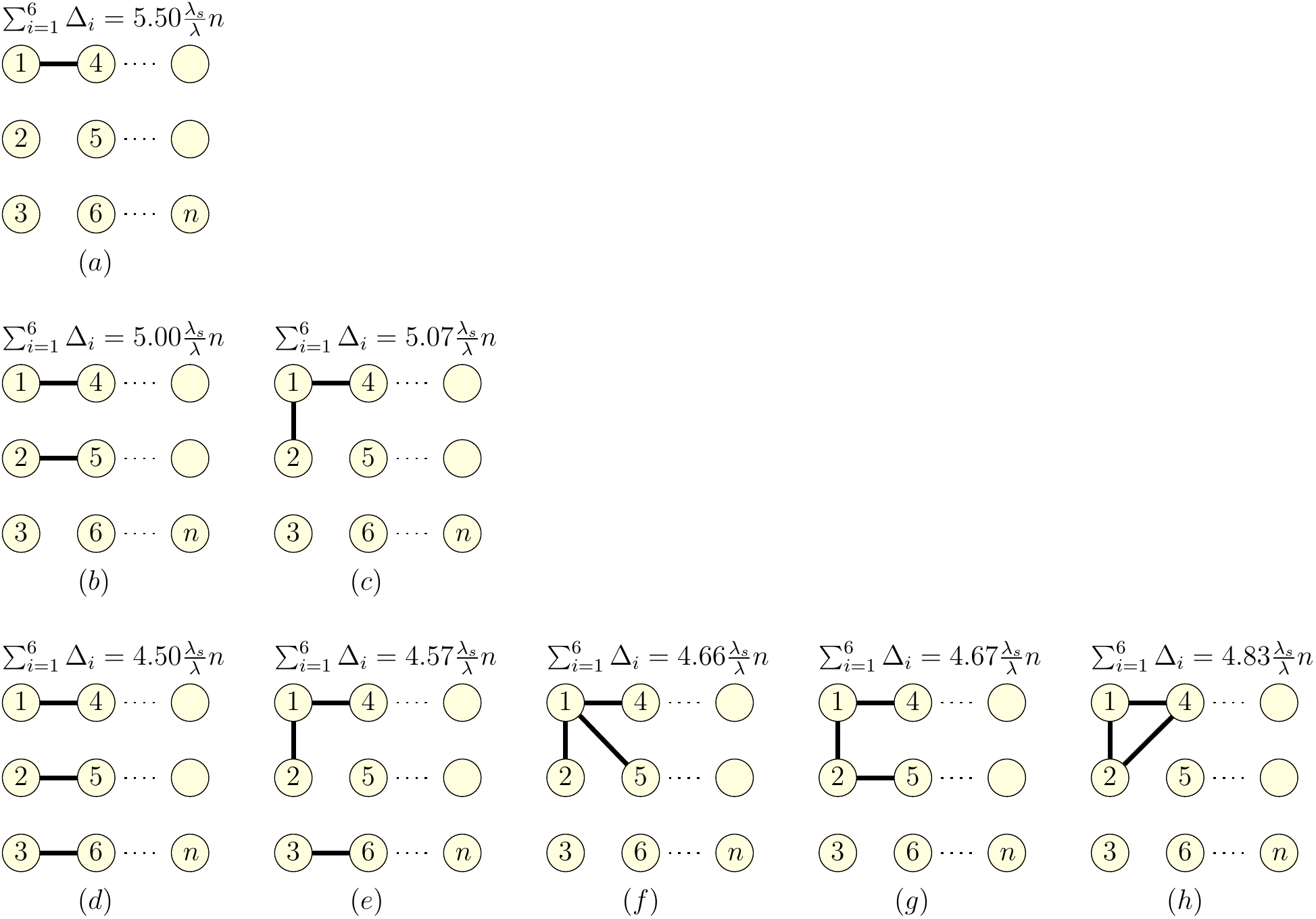}}
\caption{A network of $n$ isolated nodes, where all unique network configurations along with their corresponding values of $\sum_{i=1}^{6}\Delta_i$ are presented when $\Bar{n}$ links are added to this network, such that, $\Bar{n}=1$ in (a), $\Bar{n}=2$ in (b), (c), and $\Bar{n}=3$ in (d), (e), (f), (g), (h). Links get consolidated and number of nodes engaged decreases from left to right in each row.}
\label{fig:1_2_3_links}
\vspace*{-0.4cm}
\end{figure}

A fully connected network in the absence of any adversary has $\binom{n}{2}=\frac{n(n-1)}{2}$ links. Therefore, choosing positions of $\Tilde{n}$ jammers in a fully connected network is the same as placing $\Bar{n}=\binom{n}{2}-\Tilde{n}$ links of rate $\frac{\lambda}{n}$ in a group of $n$ isolated nodes, as in Fig.~\ref{fig:1_2_3_links}.

\begin{figure}[t]
\centerline{\includegraphics[width=0.6\linewidth]{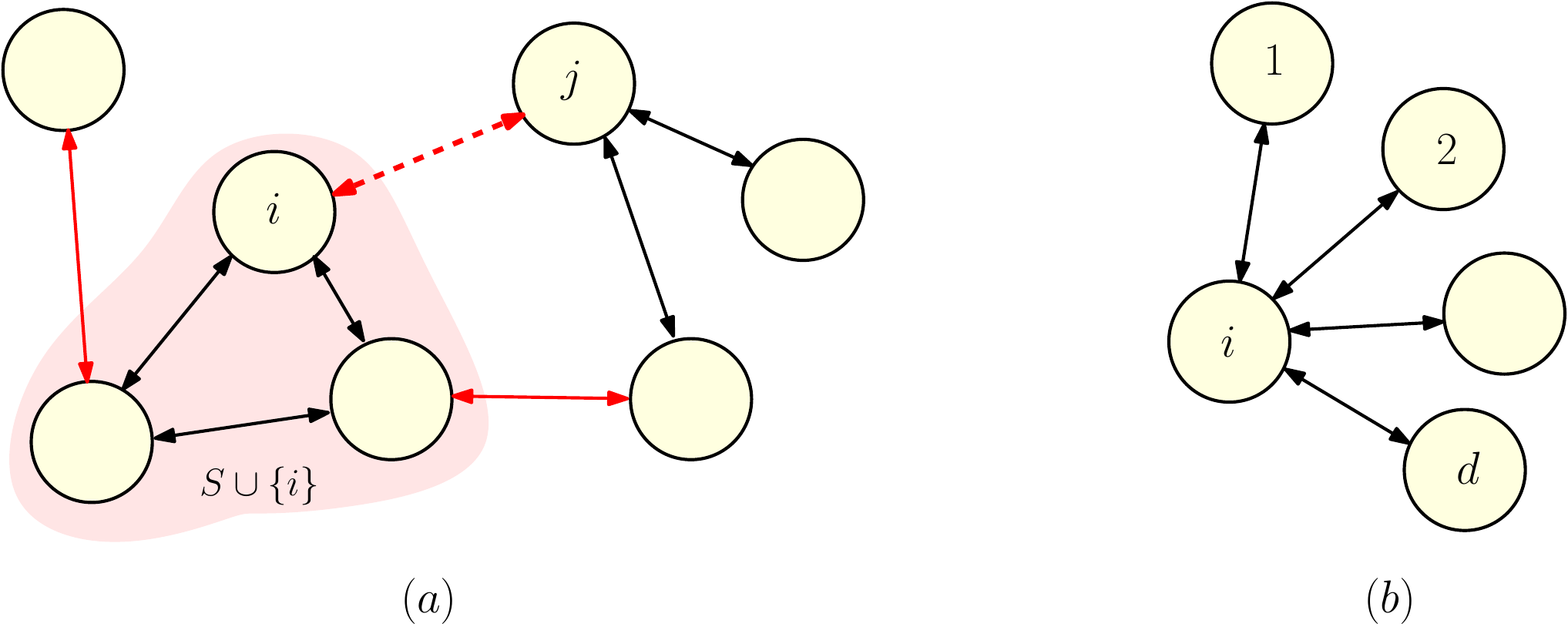}}
\caption{ (a) A new link (dotted line) is being added between node $i$ and node $j$, such that it increases the number of incoming link to the set $S\cup\{i\}$ (in orange background) by one. (b) Node $i$ of degree $d$, connected with $d$ nodes of degree $1$.} 
\label{fig:d_degree}
\vspace*{-0.4cm}
\end{figure}

\begin{lemma} \label{lemma:adding_links}
Consider an arbitrarily connected network of $n$ nodes, where the long-term expected version age at node $\ell$ is $\Delta_{\ell}$. If a new link is introduced in this network, then the long-term expected version age at node $\ell$ in the resultant network $\Bar{\Delta}_{\ell}$ satisfies $\Bar{\Delta}_{\ell} \leq \Delta_{\ell}$, for all ${\ell}$.
\end{lemma}

\begin{Proof}
Heuristically, the presence of an additional link offers the network the opportunity to have more age-conformed transitions through this link, which would lead to improved age at the nodes. Mathematically, assume that the additional link is introduced between the previously disconnected nodes $i$ and $j$, as shown in Fig.~\ref{fig:d_degree}(a), and consider a set $S_1 \subseteq \mathcal{N}\backslash \{i,j\}$. Then, for sets of the form $S=S_1 \cup \{i,j\}$, we will have $\Bar{\Delta}_S=\Delta_S$, since repeated application of (\ref{eqn:yates_eqn}) yields identical set of recursive equations for both $\Bar{\Delta}_S$ and $\Delta_S$. 
    
However, for sets of the form $S=S_1\cup \{i\}$, we will have $\Bar{\Delta}_S\leq\Delta_S$. This is because the additional link $(i,j)$ introduces additional terms of $\lambda_{ij}\Bar{\Delta}_{S\cup \{j\}}$ and $\lambda_{ij}$ in the numerator and denominator, respectively, of the equation for $\Bar{\Delta}_S$ obtained from applying (\ref{eqn:yates_eqn}), compared to $\Delta_S$. Since $S\subset S\cup \{j\}$, re-employing the trick of (\ref{eqn:k-1_mirrorblock_age_trans})-(\ref{eqn:k-1_mirrorblock_age_trans_inequality}), we use the inequality $\Bar{\Delta}_{S\cup \{j\}}\leq \Bar{\Delta}_S$ in the numerator to get rid of $\Bar{\Delta}_{S\cup \{j\}}$ term and consequently $\Bar{\Delta}_S\leq \Delta_S$ can be verified again by the set of recursive equations for $\Bar{\Delta}_S$ and $\Delta_S$. Similarly, we can show for sets of the form $S=S_1\cup \{j\}$, we will have $\Bar{\Delta}_S\leq\Delta_S$. Finally, a single application of (\ref{eqn:yates_eqn}) for the sets of the form $S=S_1$ yields similar equation for both $\Bar{\Delta}_{S}$ and $\Delta_{S}$, as $S_1$ will have same neighbors in both the networks.

Hence, for any set of nodes $S$, $\Bar{\Delta}_S\leq\Delta_S$. The lemma follows from choosing $S=\{\ell\}$, $\ell \in \mathcal{N}$, such that there exists a path from node $\ell$ to either node $i$ or node $j$ in the original network. If no such path exists, then $\Bar{\Delta}_{\ell}=\Delta_{\ell}$, as the corresponding set of recursive equations are identical in both networks and independent of the age processes at node $i$ or node $j$.
\end{Proof}

We define the degree of a node henceforth as the number of network nodes (excluding the source node) connected to it. An interesting outcome of Lemma~\ref{lemma:adding_links} is that the maximum possible age at a typical node $i$ with degree $d$ is obtained when it is connected to $d$ nodes of degree $1$, as shown in Fig.~\ref{fig:d_degree}(b). This is because, deleting any link in Fig.~\ref{fig:d_degree}(b) would cause node $i$ not to have degree $d$ anymore, while adding more links would lower the age at the node $i$ due to Lemma~\ref{lemma:adding_links}. In this case, to compute the age $\Delta_i$ at node $i$, note that the age processes at all the nodes in the set $\{1,2,\ldots,d\}$ have statistically similar age processes, and therefore, without loss of generality, we will represent any subset of $d_1$ nodes in this set, for $d_1\leq d$, by $\{1,\ldots,d_1\}$. In this case, $\Delta_{\{i\}\cup\{1,\ldots,d_1\}}$ can be characterized using (\ref{eqn:yates_eqn}) as
\begin{align} \label{eqn:d_deg_itera}
    \Delta_{\{i\}\cup\{1,\ldots,d_1\}} &= \frac{\lambda_s+\frac{(d-d_1)\lambda }{n}\Delta_{\{i\}\cup\{1,\ldots,d_1+1\}}}{\frac{(d_1+1)\lambda}{n}+\frac{(d-d_1)\lambda}{n}} \\
    &= \frac{\lambda_s+\frac{(d-d_1)\lambda }{n}\Delta_{\{i\}\cup\{1,\ldots,d_1+1\}}}{\frac{(d+1)\lambda}{n}}
\end{align}
where by iterative substitution of $d_1=d, d-1,\ldots, 0$, finally gives
\begin{align} \label{eqn:d_deg_fullexp}
    \Delta_i=\frac{\lambda_s}{\lambda}\frac{n}{(d+1)}\left(1+ \sum_{d_2=0}^{d-1}\prod_{d_1=0}^{d_2}\frac{d-d_1}{d+1}\right)
\end{align}

Note that if node $i$ was isolated, meaning it had degree zero whereby it was not connected to any other network node and only received packets from the source with rate $\frac{\lambda}{n}$, then its age would be $\frac{\lambda_s}{\lambda}n$. This implies that if $d$ links are added to an isolated node, then the age reduction at the node is lower bounded by $R_d$, where
\begin{align}
    R_d&=\left[\frac{\lambda_s}{\lambda}n \right] -\left[ \frac{\lambda_s}{\lambda}\frac{n}{(d+1)}\left(1+ \sum_{d_2=0}^{d-1}\prod_{d_1=0}^{d_2}\frac{d-d_1}{d+1}\right) \right] \\
    &= \frac{\lambda_s}{\lambda}n \left[1 - \frac{1}{(d+1)}\left(1+ \sum_{d_2=0}^{d-1}\prod_{d_1=0}^{d_2}\frac{d-d_1}{d+1}\right) \right] \label{eqn:Rd_exp}
\end{align}

Table~\ref{table_Rd_values} below lists coefficients of $\frac{\lambda_s}{\lambda}n$ in $R_d$ values for $d=1,\ldots,22$, where $R_d/\frac{\lambda_s}{\lambda}n $ has been rounded down up to two decimal points, thereby giving a further lower bound.

\begin{table}[h]
\begin{center}
 \begin{tabular}{|l | l | l | l| l| l| l| l| l| l| l| l|} 
 \hline
  $d$ & $1$ & $2$ & $3$ & $4$ & $5$ & $6$ & $7$ & $8$ & $9$ & $10$ & $11$ \\
 \hline 
 $R_d/\frac{\lambda_s}{\lambda}n$ & $0.25$ & $0.37$ & $0.44$ & $0.49$ & $0.53$ & $0.56$ & $0.59$ & $0.61$ & $0.63$ & $0.64$ & 0.66\\
 \hline
 \end{tabular} 
 
 \vspace{0.5em}
 
 \begin{tabular}{|l | l | l | l| l| l| l| l| l| l| l| l|} 
  \hline
  $d$ & $12$ & $13$ & $14$ & $15$ & $16$ & $17$ & $18$ & $19$ & $20$ & $21$ & $22$\\
 \hline 
 $R_d/\frac{\lambda_s}{\lambda}n$ & $0.67$ & $0.68$ & $0.69$ & $0.70$ & $0.71$ & $0.72$ & $0.72$ & $0.73$ & $0.74$ & $0.74$ & $0.75$ \\
 \hline
 \end{tabular}
\end{center}
\caption{$R_d/\frac{\lambda_s}{\lambda}n$ values (rounded down to nearest 2 decimal point number) for $d=1,\ldots, 22$.}
\label{table_Rd_values}
\vspace*{-0.5cm}
\end{table}

Notice in Table~\ref{table_Rd_values}, how adding the first link causes an age drop of $0.25\frac{\lambda_s}{\lambda}n$, but for larger $d$, adding additional links does not cause much reduction in the age of a node. Thus, adding a new link to $d$ degree node might cause less system age drop instead of adding it to an isolated node.

Given all this machinery, we seek the configuration of $\Bar{n}$ links on a group of $n$ isolated nodes that would result in the highest total age (or average age) for the network. Fig.~\ref{fig:1_2_3_links} shows all possible configurations with number of links $\Bar{n}=1,2,3$, along with the exact evaluation of their total age for the network. Since $\Bar{n}\leq 3$ links can engage with at most 6 distinct nodes, we focus on the total age of the set of nodes $\{1,2,\ldots,6\}$, as the remaining (isolated) nodes have the same total age of $\sum_{i=7}^{n}\Delta_i=(n-6)\frac{\lambda_s}{\lambda}n$ for all configurations of Fig.~\ref{fig:1_2_3_links}. Here $\Delta_i$, corresponding to the expected age of node $i$, is computed through the set of recursive equations obtained from (\ref{eqn:yates_eqn}). For example, in the case of Fig.~\ref{fig:1_2_3_links}(a), these recursive equations for nodes $1$ and $4$ are
\begin{align}
    \Delta_{\{1,4\}}=0.5\frac{\lambda_s}{\lambda}n \quad \Rightarrow \quad
    \Delta_1=\Delta_4=0.5\frac{\lambda_s}{\lambda}n+0.5\Delta_{\{1,4\}}=0.75\frac{\lambda_s}{\lambda}n 
\end{align}
which give the age of nodes $1$ and $4$ each as $0.75\frac{\lambda_s}{\lambda}n$. For isolated nodes $2$, $3$, $5$, we have
\begin{align}
    \Delta_2=\Delta_3=\Delta_5=\Delta_6=\frac{\lambda_s}{\lambda}n
\end{align}
Thus, the total age for $\{1,2,\ldots,6\}$ in this case is $(0.75+0.75+1+1+1+1)\frac{\lambda_s}{\lambda}n=5.50 \frac{\lambda_s}{\lambda}n$.

In each row of Fig.~\ref{fig:1_2_3_links}, all network configurations have the same number of links, such that the number of nodes engaged by these links decreases from left to right. For example in the third row, the left most configuration engages six nodes, making their degree one, and the right most network engages three nodes, fully interconnected amongst themselves, forming a mini-fully connected network (mini-FC) of size three. We observe in each row that the total age of the network increases from left to right with consolidation of the links to fewer engaged nodes, leaving higher number of nodes isolated. We observed something similar in Section~\ref{subsec:position_jammers} for the ring network, where the worst configuration involved consolidating all links in one single line network thereby creating multiple isolated nodes, whereas the most favorable configuration involved spreading the links more evenly around the ring. However, in general for $n$-node network with $\bar{n}$ links, there are $\binom{\binom{n}{2}}{\Bar{n}}=O(n^2/\Bar{n})$ different possible network configurations, and unlike a ring network where the dismembered components always exhibited line topology, very complicated topologies can arise in an arbitrarily jammed fully connected network. This makes it difficult to pick out the worst configuration due to the nature of recursive equations which may require computation of ages of up to $2^n$ unique sets of nodes for deriving the age at all network nodes. Since finding the optimum configuration is exponentially complex, next we develop a greedy approach of placing $\Bar{n}$ links with the goal of maximizing the average age in the resultant network.

\subsection{Greedy Approach}\label{subsec:greedy}

In our greedy approach, instead of placing all $\Bar{n}$ links in one go, we place them in $\Bar{k}$ steps such that $\binom{\Bar{k}-1}{2} \leq \Bar{n} \leq \binom{\Bar{k}}{2}$. At step $k$, we have a network of $\binom{k}{2}$ links, in which we place $k$ new links in a manner that maximizes the total age of the resultant network of $\binom{k+1}{2}$ links. In Lemma~\ref{lemma:greedy} next, mini-FC (mini fully connected) of $k$ nodes corresponds to set of $k \leq n$ nodes where every node receives update packets from the source with rate $\frac{\lambda}{n}$ and from each of the other $k-1$ nodes with link of rate $\frac{\lambda}{n}$, such that the $k$ nodes are fully connected amongst themselves.

\begin{lemma}\label{lemma:greedy}
    Consider a network of $n$ nodes where $k$ nodes are connected in a mini-FC and $n-k$ nodes are isolated. Given that we place $k$ new links in this network, the resultant network will have the maximum total network age when the new $k$ links are placed such that they connect a single isolated node to all the $k$ nodes of the mini-FC, resulting in a network where $k+1$ nodes are connected in a mini-FC and $n-k-1$ nodes are isolated.
\end{lemma}

\begin{Proof}
In a mini-FC of $k$ nodes, the age at each node, using (\ref{eqn:yates_eqn}), is $\frac{\lambda_s}{\lambda}\frac{n}{k}\left(\sum_{j=1}^{k}\frac{1}{j}\right)$. Then, the total age in network of $n$ nodes with $k$ nodes connected in mini-FC and $n\!-\!k$ nodes isolated is
\begin{align}
    \sum_{i=1}^{n}\Delta_i &= k \times \frac{\lambda_s}{\lambda}\frac{n}{k}\left(\sum_{j=1}^{k}\frac{1}{j}\right) + (n-k) \times \frac{\lambda_s}{\lambda}n \\
    &= \frac{\lambda_s}{\lambda} \left(  n\left(\sum_{j=1}^{k}\frac{1}{j}\right) + n(n-k) \right) \label{eqn:totalage_miniFC_k}
\end{align}
If $k$ new links are added to this network in a manner that connects an isolated node to the $k$ nodes of this mini-FC, as in Fig.~\ref{fig:engaging_1_node}(a), the total age of the resultant network will be
\begin{align} 
    \sum_{i=1}^{n}\Delta_i &= (k+1) \times \frac{\lambda_s}{\lambda}\frac{n}{k+1}\left(\sum_{j=1}^{k+1}\frac{1}{j}\right) + (n-k-1) \times \frac{\lambda_s}{\lambda}n \\
    &= \frac{\lambda_s}{\lambda} \left(  n\left(\sum_{j=1}^{k+1}\frac{1}{j}\right) + n(n-k-1) \right) \label{eqn:totalage_miniFC_k_1}
\end{align}
As a consequence of adding these $k$ links, the total age of the system reduces by
\begin{align}
    \Bar{R} &=\frac{\lambda_s}{\lambda} \left(  n\left(\sum_{j=1}^{k}\frac{1}{j}\right) + n(n-k) \right) - \frac{\lambda_s}{\lambda} \left(  n\left(\sum_{j=1}^{k+1}\frac{1}{j}\right) + n(n-k-1) \right) \\
    &=\frac{\lambda_s}{\lambda} \left( n-n\frac{1}{k+1} \right) \\
    &=\frac{\lambda_s}{\lambda}n \left( \frac{k}{k+1} \right)\label{eqn:bar_R}
\end{align}

We will now show that if the $k$ links were instead added in any other manner, the total age of the network would drop by more than $\Bar{R}$. Thus, starting with a $k$ node mini-FC and $n-k$ isolated nodes, the worst (maximum age) configuration that can be obtained is a $k+1$ node mini-FC and $n-k-1$ isolated nodes, when $k$ new links are added. 

\begin{figure}[t]
\centerline{\includegraphics[width=0.9\linewidth]{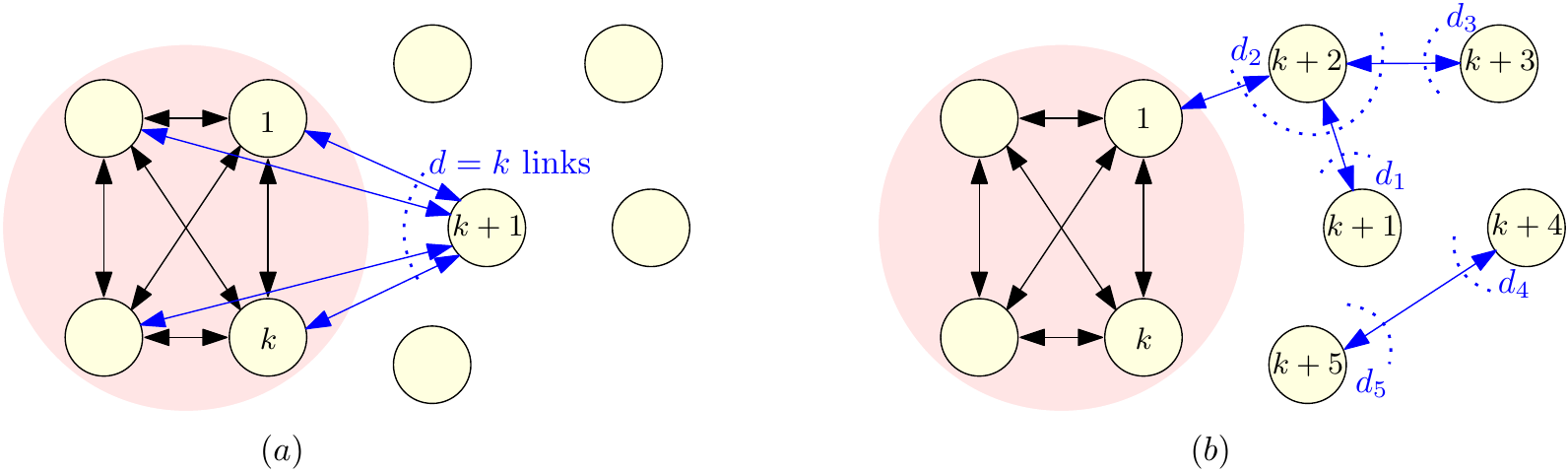}}
\caption{$k=4$ links are added to a network size $k=4$ mini-FC and $n-k=5$ isolated nodes. (a) All $k$ links are placed such that they connect node $k+1$ to all nodes of mini-FC, thereby creating a network of size $k+1=5$ mini-FC and $n-k=4$ isolated nodes. The degree of node $k+1$ is $d=k$. (b) The additional $k$ links engage multiple nodes, such that their sum of degrees $d_1+d_2+d_3+d_4=7\geq k$.}
\label{fig:engaging_1_node}
\vspace*{-0.4cm}
\end{figure}

Without loss of generality, let the nodes of the mini-FC correspond to the set of node indices $\{1,\ldots,k\}$. Note that no new link can be placed between any two nodes of this set, since all links in the mini-FC are already saturated. Hence, each new link will engage at least one node in the set $\{k+1,\ldots,n\}$ of previously isolated nodes. If all the $k$ links were to engage exactly one isolated node, say node $k+1$, that would be possible only if the $k$ links connect the single node $k+1$ to all the nodes $\{1,\ldots,k\}$ of the mini-FC, which just results in a network of $k+1$-size mini-FC and $n-k-1$ isolated nodes as described above. Hence, to prove Lemma~\ref{lemma:greedy}, it suffices to show that engaging two or more nodes from the set $\{k+1,\ldots,n\}$ is not optimal.

It is easy to see that engaging four or more nodes is not optimal. This is because from Table~\ref{table_Rd_values}, $R_1=0.25\frac{\lambda_s}{\lambda}n$, and from Lemma~\ref{lemma:adding_links}, $R_d$ will be be an increasing function of $d$, as a higher degree node has more links which reduces the node age. Since engaging a node means making at least one link incident on it, engaging four or more nodes would give a minimum system age reduction of $4\times R_1=\frac{\lambda_s}{\lambda}n$. This age reduction is more than (\ref{eqn:bar_R}), since $\Bar{R} \leq \frac{\lambda_s}{\lambda}n$.

Next, we show that engaging three nodes is not optimal. Let the three engaged nodes be $k+1$, $k+2$ and $k+3$, such that their degrees after placing the $k$ links become $d_1$, $d_2$ and $d_3$. Note that $d_1+d_2+d_3 \geq k$, where the equality holds when each new link connects a node from set $\{k+1, k+2, k+3\}$ to a node in the set $\{1,\ldots,k\}$. The inequality comes from the fact that when a new link connects, say, node $k+1$ and node $k+2$, it increments both degrees $d_1$ and $d_2$ by one, hence each such link gets counted twice, in degrees of both nodes, as shown in Fig.~\ref{fig:engaging_1_node}(b). 

Consider the case of $k=7$, where Table~\ref{table_7links} lists all possible unique ways of satisfying $d_1+d_2+d_3 = 7$ and $R_{d_1}+R_{d_2}+R_{d_3}$ represents the minimum age reduction caused by turning three previously isolated nodes into nodes of degree $d_1$, $d_2$ and $d_3$. We can see in Table~\ref{table_7links} that the system age reduction caused by engaging three nodes will be greater $\frac{\lambda_s}{\lambda}n>\Bar{R}$ in all configurations, when $k=7$ and $d_1+d_2+d_3 = 7$. Note that the age reduction values in Table~\ref{table_7links} are a lower bound to the actual system age reduction, since we are only considering a lower bound to the age reduction caused at three nodes of degree $d_1$, $d_2$ and $d_3$, and we have not accounted for the age reduction at nodes of the set $\{1,\ldots,k\}$ due to the new links. 

\begin{table}[h]
\begin{center}
\begin{tabular}{|l | l | l | l| l| l| l| l| l| l| l|} 
 \hline
  $d_1$ & $d_2$ & $d_3$ & $(R_{d_1}+R_{d_2}+R_{d_3})/\frac{\lambda_s}{\lambda}n$ \\
 \hline 
 $5$  & $1$ & $1$ & $0.53+0.25+0.25=1.03$   \\
 \hline 
 $4$  & $2$ & $1$ & $0.49+0.37+0.25=1.11$   \\
 \hline 
 $3$  & $3$ & $1$ &  $0.44+0.44+0.37=1.13$ \\
 \hline 
 $3$  & $2$ & $2$ & $0.44+0.37+0.37=1.18$   \\
 \hline
\end{tabular}
\end{center}
\caption{Minimum age reduction with 3 node engagement and 7 additional links.}
\label{table_7links}
\vspace*{-0.4cm}
\end{table}

If $d_1+d_2+d_3 > 7$, then the age reduction caused is even higher than the values of Table~\ref{table_7links}, since $R_d$ is an increasing function of $d$. Likewise $k>7$ implies $d_1+d_2+d_3 \geq k >7$, and hence, age reduction is again higher than $\Bar{R}$. Finally, for $d_1+d_2+d_3 < 7$, we have numerically verified that the age reduction is always more than $\Bar{R} =\frac{\lambda_s}{\lambda}n \left( \frac{k}{k+1} \right)$. For example, consider $(d_1,d_2,d_3)=(2,1,1)$, in which case $k \leq 4$ and $\Bar{R}\leq \frac{4}{5}\frac{\lambda_s}{\lambda}n $ since $\Bar{R}$ is an increasing function of $k$. From Table~\ref{table_Rd_values}, in this case $R_{d_1}+R_{d_2}+R_{d_3}=(0.37+0.25+0.25)\frac{\lambda_s}{\lambda}n=0.87\frac{\lambda_s}{\lambda}n \geq \frac{4}{5}\frac{\lambda_s}{\lambda}n \geq \Bar{R}$. 

Next, we show that engaging two nodes is not optimal. In this case Table~\ref{table_23links} lists all possible pairs $(d_1,d_2)$ satisfying $d_1+d_2=23$ and the corresponding minimum age reduction caused by two nodes of degree $d_1$ and $d_2$, where we see $R_{d_1}+R_{d_2} \geq \frac{\lambda_s}{\lambda}n >\Bar{R}$ for all listed pairs $(d_1,d_2)$. Hence, for $d_1+d_2 \geq23$, we have a minimum system age reduction of $\frac{\lambda_s}{\lambda}n >\Bar{R}$. Further, for all $d_1+d_2 < 23$, we have numerically verified for all possible pairs that the system age reduction is higher than $\Bar{R}$, similar to the three node case. Observe how in Table~\ref{table_7links} and Table~\ref{table_23links}, when links are more concentrated in $d_1$, the lower bound for age reduction is poorer, but as links get more evenly distributed amongst the engaged nodes, the age reduction lower bound increases, which is in line with the general worstness of link consolidation for the system age observed in ring networks in Section~\ref{subsec:position_jammers}, and for fully connected networks for $\Bar{n}\leq 3$ in Fig.~\ref{fig:1_2_3_links}.

\begin{table}[h]
\begin{center}
\begin{tabular}{|l | l | l |} 
 \hline
  $d_1$ & $d_2$ & $(R_{d_1}+R_{d_2})/\frac{\lambda_s}{\lambda}n$ \\
 \hline 
 $22$  & $1$ &  $0.75+0.25=1.00$  \\
 \hline 
 $21$  & $2$ &  $0.74+0.37=1.11$  \\
 \hline 
 $20$  & $3$ &  $0.74+0.44= 1.18$ \\
 \hline 
 $19$  & $4$ &   $0.73+0.49= 1.22$ \\
 \hline
 $18$  & $5$ &   $0.72+0.53= 1.25$ \\
 \hline
 $17$  & $6$ &   $0.72+0.56= 1.28$ \\
 \hline
 $16$  & $7$ &   $0.71+0.59= 1.30$ \\
 \hline
 $15$  & $8$ &   $0.70+0.61= 1.31$ \\
 \hline
 $14$  & $9$ &   $0.69+0.63= 1.32$ \\
 \hline
 $13$  & $10$ &  $0.68+0.64= 1.32$ \\
 \hline
 $12$  & $11$ &  $0.67+0.66= 1.33$ \\
 \hline
\end{tabular}
\end{center}
\caption{Minimum age reduction with 2 node engagement and 23 additional links.}
\label{table_23links}
\vspace*{-0.4cm}
\end{table}

Hence, engaging just one node yields the minimum possible age reduction, which upon adding $k$ links to the network of size $k$ mini-FC and $n-k$ isolated nodes results in a network of size $k+1$ mini-FC and $n-k-1$ isolated nodes.
\end{Proof}

Going back to our greedy approach, at step 1, we add one link to a network of $n$ isolated nodes, that creates a size-2 mini-FC and $n-2$ isolated nodes. Next, at step 2, the greedy approach adds two links with the goal of maximizing the age of the resultant system of three links. From Lemma~\ref{lemma:greedy}, we know that greedy approach will engage just one additional node, resulting in a size-3 mini-FC, as in Fig.~\ref{fig:1_2_3_links}(h). Similarly in later steps, the greedy approach will keep adding links in consolidated manner that engages no more than one node at every step, leaving majority of nodes isolated. At the last step, we just add the remaining links again to a single new node.

\begin{figure}[t]
\centerline{\includegraphics[width=0.9\linewidth]{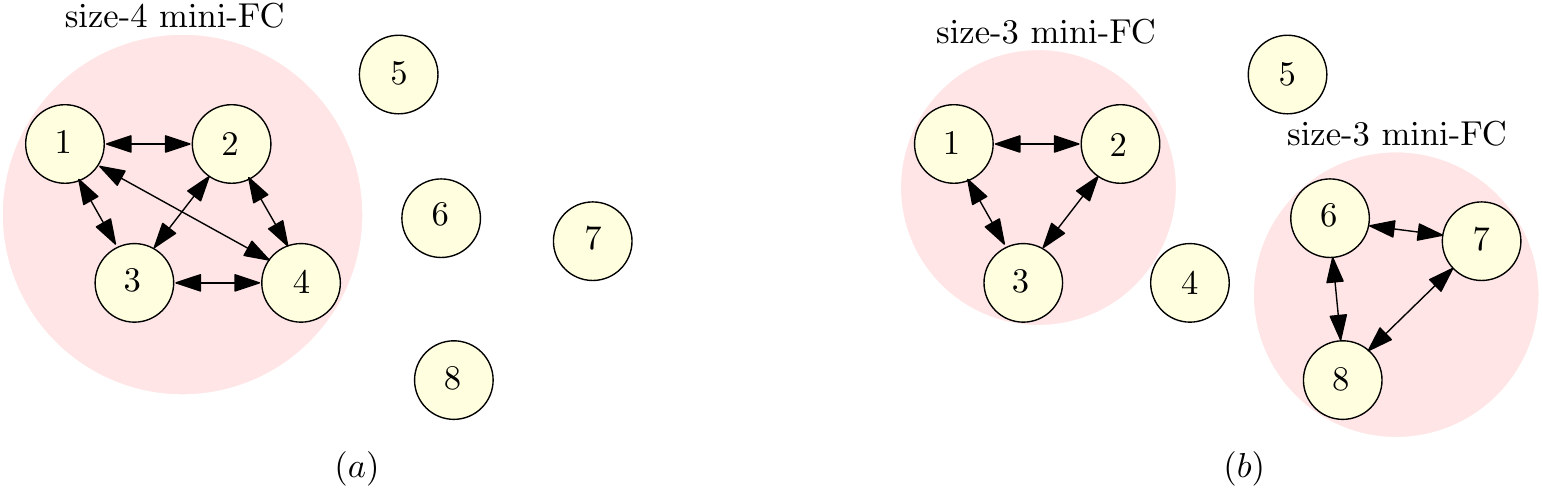}}
\caption{(a) A network with $\Bar{n}=6$ links, consisting of one size $k=4$ mini-FC and $4$ isolated nodes. (b) A network with $\Bar{n}=6$ links, consisting of $\Bar{m}=2$ clusters of mini-FCs each of size $\Bar{k}=3$ and $2$ isolated nodes.}
\label{fig:clustering_miniFC}
\vspace*{-0.4cm}
\end{figure}

However, it is not clear if the greedy approach gives the worst network configuration over all possible network configurations with $\Bar{n}=\binom{k+1}{2}$ links, $k \in \{1,\ldots,n-1\}$, since Lemma~\ref{lemma:greedy} only compares all networks that have at least one size-$k$ mini FC and a total of $\binom{k+1}{2}$ links, which is a subset of networks that can be created with $\binom{k+1}{2}$ links. Though it is difficult to compare all network configurations, it would be interesting to see if link placement by the greedy approach, which creates one single mini-FC of size $k$, also results in higher total age than a network containing multiple mini-FCs of smaller size with the same number of links; see Fig.~\ref{fig:clustering_miniFC}.

We know that the total network age for network with size $k$ mini-FC and $n-k$ isolated nodes is given by (\ref{eqn:totalage_miniFC_k}). Let us say that the links are instead arranged into a $\Bar{m}$ clusters or mini-FCs of size $\Bar{k}$, $\Bar{m}>1$, such that the total numbers of links are the same in both cases,
\begin{align}\label{eqn:mbar_relation}
    \Bar{m}\binom{\Bar{k}}{2}=\binom{k}{2} \quad \Rightarrow \quad
    \frac{k(k-1)}{\Bar{k}(\Bar{k}-1)}=\Bar{m}>1
\end{align}
Since the age at each node in a mini-FC of $\Bar{k}$ nodes is $\frac{\lambda_s}{\lambda}\frac{n}{\Bar{k}}\left(\sum_{j=1}^{\Bar{k}}\frac{1}{j}\right)$, the total network age in the second network configuration, e.g., see Fig.~\ref{fig:clustering_miniFC}(b), would be
\begin{align} 
    \frac{\lambda_s}{\lambda}\left( \Bar{m}\Bar{k}\frac{n}{\Bar{k}}\left( \sum_{j=1}^{\Bar{k}} \frac{1}{j}\right) + (n-\Bar{m}\Bar{k})n \right) 
    &= \frac{\lambda_s}{\lambda}\left( \frac{k(k-1)}{\Bar{k}(\Bar{k}-1)}n\left( \sum_{j=1}^{\Bar{k}} \frac{1}{j}\right) + (n-\frac{k(k-1)}{\Bar{k}(\Bar{k}-1)}\Bar{k})n \right) \\
    &= \frac{\lambda_s}{\lambda}\left( \frac{k(k-1)}{\Bar{k}(\Bar{k}-1)}n\left(- \sum_{j=1}^{\Bar{k}} \frac{j-1}{j}\right) + n^2 \right) \\
    & \leq \frac{\lambda_s}{\lambda}\left( n\left(- \sum_{j=1}^{k} \frac{j-1}{j}\right) + n^2 \right) \label{eqn:clustering_vs_single_1} \\
    &= \frac{\lambda_s}{\lambda}\left( n\left( \sum_{j=1}^{k} \frac{1}{j}\right) + (n-k)n \right) \label{eqn:clustering_vs_single}
\end{align}
where the inequality in (\ref{eqn:clustering_vs_single_1}) comes from $\frac{1}{\Bar{k}(\Bar{k}-1)}\left( \sum_{j=1}^{\Bar{k}} \frac{j-1}{j}\right)$ being a decreasing function of $\Bar{k}$, which in turn can be verified by comparing the expression for $\Bar{k}$ and $\Bar{k}+1$, and the expression in (\ref{eqn:clustering_vs_single}) is the same as (\ref{eqn:totalage_miniFC_k}). This shows that consolidation of links into one big mini-FC results in higher total network age instead of grouping the links into multiple smaller mini-FCs.

Next, we examine the robustness of the average age $\Delta=\frac{\sum_{i=1}^{n}\Delta_i}{n}$ of the network that results from the remaining $\binom{n}{2}-\Tilde{n}$ links (after placing the $\Tilde{n}$ jammers) arranged in a manner that creates the network suggested by the greedy approach.

\begin{lemma}\label{lemma:FC_nlogn}
    For a fully connected network of $n$ nodes with $\Tilde{n}$ jammers placed such that the remaining links are consolidated in accordance with the greedy approach, the network continues to have $\Delta=O(\log{n})$ so long as $\Tilde{n}=O(n\log{n})$.
\end{lemma}

\begin{Proof}
First, let us consider the case where $\Tilde{n}$ is such that there are exactly $\Bar{n}=\binom{k}{2}$ links left in the network. Then, from (\ref{eqn:totalage_miniFC_k}), the average age in network formed through greedy approach is 
\begin{align}\label{eqn:avgaage_greedy}
    \Delta=\frac{\sum_{i=1}^{n}\Delta_i}{n} =\frac{\lambda_s}{\lambda} \left(  \left(\sum_{j=1}^{k}\frac{1}{j}\right) + (n-k) \right) = O(\log{k})+ O(n-k)
\end{align}
where the harmonic series $\sum_{j=1}^{k}\frac{1}{j}$ exceeds the natural logarithm $\log{k}$ by Euler–Mascheroni constant as limit $k \to \infty$.

The age scaling in (\ref{eqn:avgaage_greedy}) will continue to be $O(\log{n})$ if $n-k=O(\log{n})$, i.e., $k=n-O(\log{n})$, in which case the total number of links in the network will be
\begin{align}\label{eqn:nolinks}
    \binom{k}{2}=\frac{k(k-1)}{2}=\frac{(n-O(\log{n}))(n-O(\log{n})-1)}{2}=\frac{n^2-O(n\log{n})}{2}=\binom{n}{2}-\Tilde{n}
\end{align}
i.e., the number of jammers $\Tilde{n}=O(n\log{n})$. 

Conversely, if $\Tilde{n}=O(n\log{n})$ jammers are given, then $\binom{k}{2}=\binom{n}{2}-\Tilde{n}$ yields
\begin{align}\label{eqn:k_nlogn}
    k=\sqrt{n^2-O(n\log{n})}\approx n-O(\log{n})
\end{align}
where we used $(1-x)^{\frac{1}{2}}\approx 1-\frac{x}{2}$ for small $x$ and the terms $\frac{k}{2}$, $\frac{n}{2}$ are masked under $O(n\log{n})$. Substituting (\ref{eqn:k_nlogn}) in (\ref{eqn:avgaage_greedy}) gives $\Delta=O(\log{n})$.

Next, let us say $\Tilde{n}$ is such that $\binom{k}{2}< \binom{n}{2}- \Tilde{n} <\binom{k}{2}+k=\binom{k+1}{2}$. Then, the age of this network can be sandwiched between the age of the network with $\binom{n}{2}-\binom{k+1}{2}$ jammers and $\binom{n}{2}-\binom{k}{2}$ jammers, both of which still have $O(n \log{n})$ average age.
\end{Proof}

Lemma~\ref{lemma:FC_nlogn} implies that the fully connected network is robust against $n\log{n}$ jammers when the jammers are placed according to the network resulting from the greedy approach. That is, both the original fully connected network and the jammed fully connected network have average version age of $\log n$ so long as the number of jammers is up to $n \log n$.

\begin{lemma}\label{lemma:FC_n_alpha}
    For a fully connected network of $n$ nodes with $\Tilde{n}=cn^{\alpha}$ jammers, $\alpha \in (1,2]$, placed in accordance with the greedy approach, the network has $\Delta=O(n^{\alpha-1})$. 
\end{lemma}

\begin{Proof}
The proof proceeds similar to (\ref{eqn:k_nlogn}). When $\Tilde{n}=cn^{\alpha}$, $\alpha \in (1,2)$, then similar to (\ref{eqn:nolinks}), the size of the mini-FC denoted by $k$ can be found as 
\begin{align}\label{eqn:k2_nalpha}
    k^2=n^2-O(n^{\alpha})
\end{align}
where all $o(n^{\alpha})$ terms are masked under the $O(n^{\alpha})$ term. Since in the regime of large $n$, $\frac{n^{\alpha}}{n^2}$ converges to zero, taking square root and using Taylor approximation, we obtain
\begin{align}\label{eqn:k_nalpha}
    k\approx n-O(n^{\alpha-1})
\end{align}
which, upon subsitution in (\ref{eqn:avgaage_greedy}) gives $\Delta = O(n^{\alpha-1})$. If instead $\Tilde{n}=cn^{2}$ with $c\in (0,\frac{1}{2})$, then in the regime of large $n$, we have $\frac{k^2}{2}=\frac{n^2}{2}-cn^{2}$, which gives $k=n\sqrt{1-2c}$. Hence, the second term in (\ref{eqn:avgaage_greedy}) is $O(n)$, thereby completing the proof.
\end{Proof}

\section{Numerical Results}

\begin{figure}[t]
\centerline{\includegraphics[width=0.55\linewidth]{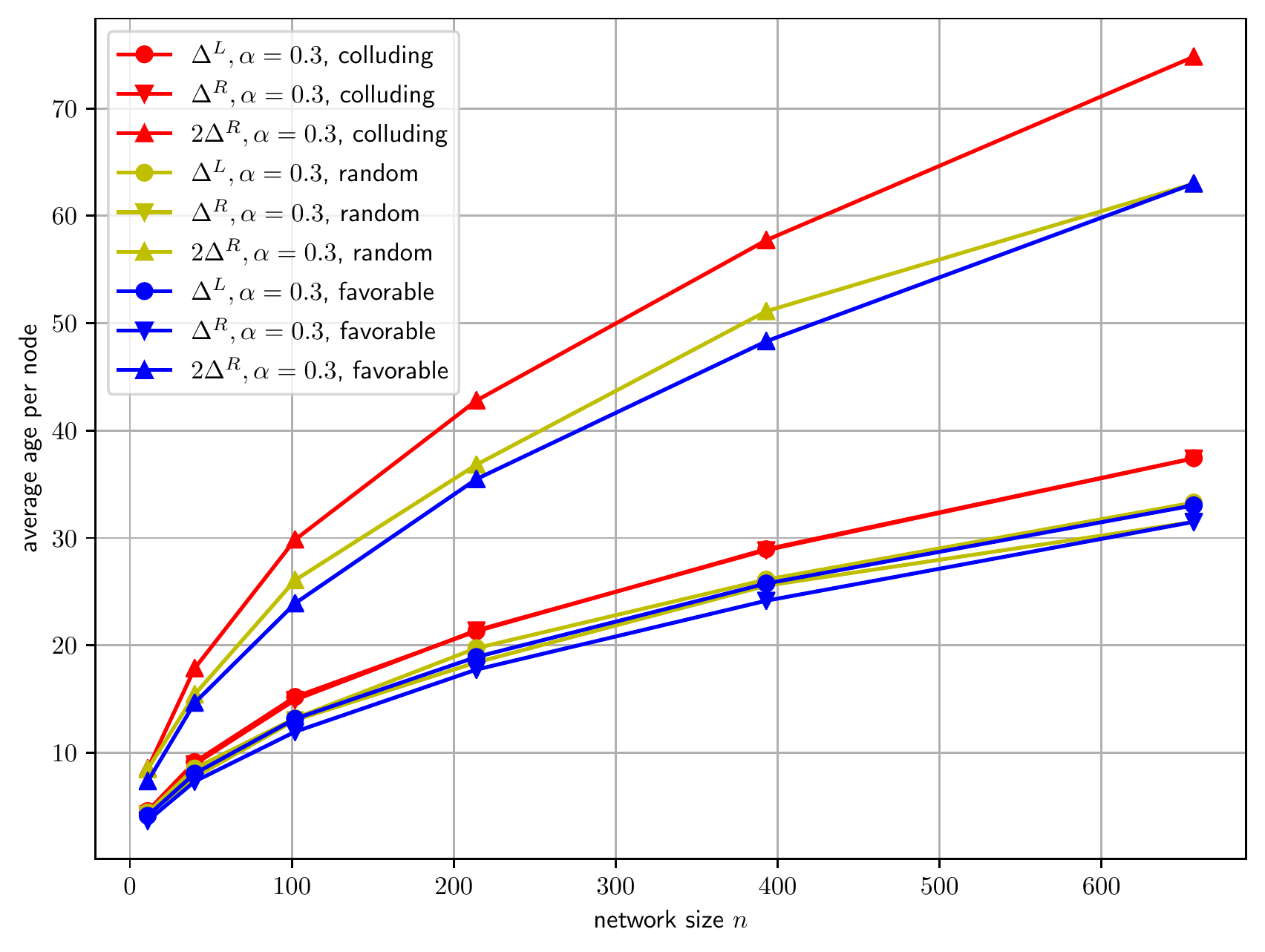}}
\caption{Average age in ring network with $n^{0.3}$ jammers}
\label{fig:ring_average_age_with_alpha3}
\end{figure}

\begin{figure}[t]
\centerline{\includegraphics[width=0.55\linewidth]{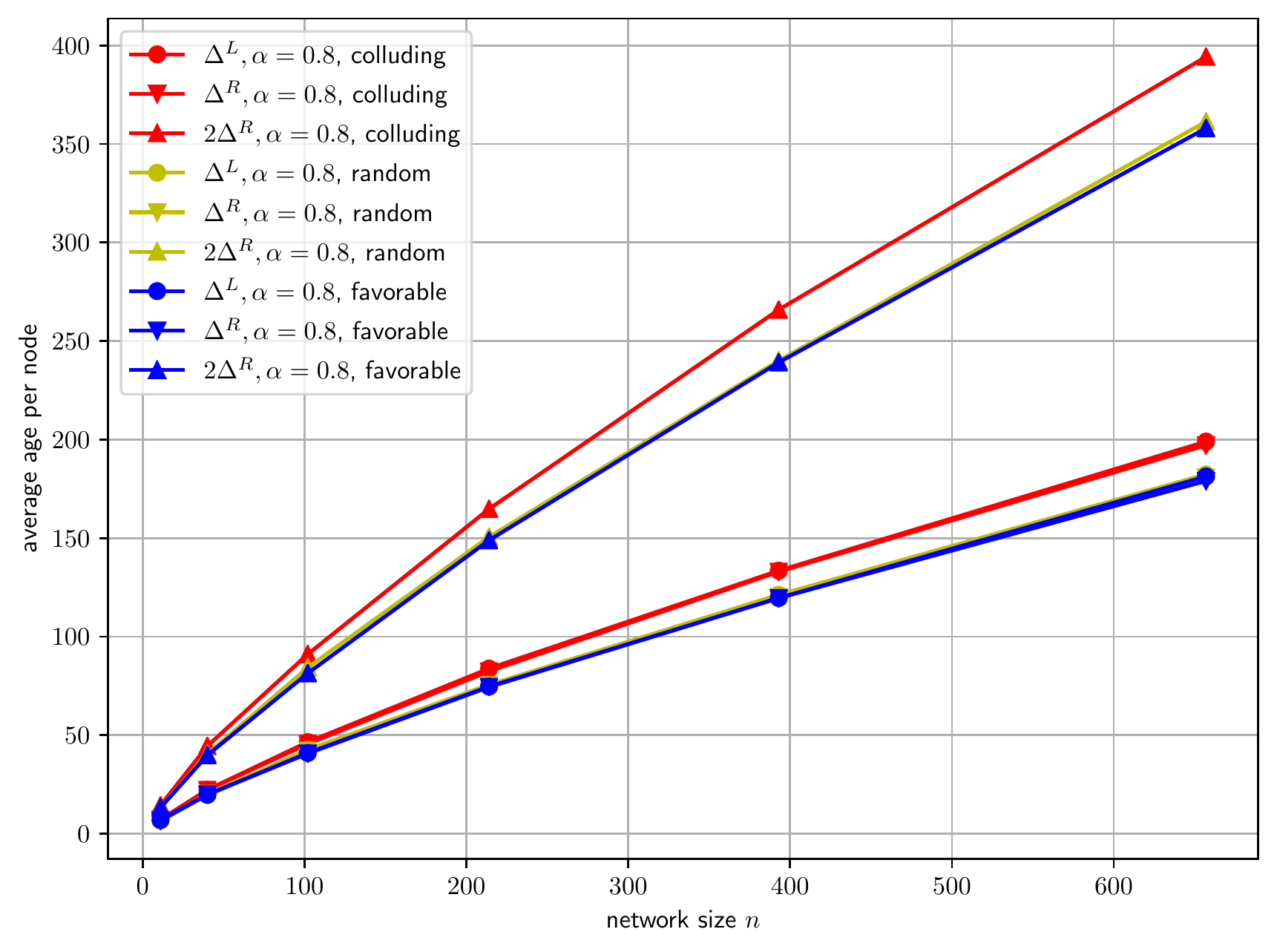}}
\caption{Average age in ring network with $n^{0.8}$ jammers.}
\label{fig:ring_average_age_with_alpha8}
\vspace*{-0.4cm}
\end{figure}

We first validate the aforementioned bounds in the case of ring networks by performing real-time gossip protocol simulations for both dismembered ring and its altered mini-ring model for three cases of jammer positions, a) most harmful or colluding jammer positions, b) random jammer positions, and c) least harmful or favorable jammer positions. For $\tilde n= n^{\alpha}$ and $\frac{\lambda_s}{\lambda}=1$, Fig.~\ref{fig:ring_average_age_with_alpha3} shows the average age plots as a function of the network size $n$ for $\alpha=0.3$. Since the number of jammers $\tilde n$ must be an integer, we choose $\tilde n = \lfloor n^{0.3} \rfloor$ and accordingly pick $n= \lceil {\tilde n}^{1/0.3} \rceil$, $\tilde n \in \mathbb{N}$ on horizontal axis to have $\tilde n$ increase consistently with $n$. In Fig.~\ref{fig:ring_average_age_with_alpha3}, we observe that $\Delta^{\ell}$ is closely approximated by its lower bound $\Delta^r$ in all three cases and both fall between $\sqrt{\frac{\pi}{2}}\sqrt{n} +e^{-\frac{n^{(1-2\times 0.3)}}{2}} n^{0.3}$ of (\ref{eqn:lowerbound_alpha3}) and $n^{0.3} + \sqrt{\frac{\pi}{2}}\sqrt{n}$ of (\ref{eqn:upperbound}). Notice that the (yellow) graph for the case of random jammer placements is irregular, since we have not chosen any persistent pattern for the choice of jammer positions, unlike the colluding and favorable positions cases. Nevertheless, the age in the former always falls between the latter two extremes. We further note that, in the case of $\alpha =0.8$ plotted in Fig.~\ref{fig:ring_average_age_with_alpha8}, where the age values increase steeply owing to the presence of larger number of jammers, the graphs look almost linear because as $\alpha$ approaches $1$, $n^{\alpha}$ begins to appear linear. Here again, $\Delta^{\ell}$ and $\Delta^r$ almost overlap, and both fall between $n^{0.8}$ of (\ref{eqn:lowerbound_alpha8}) and $n^{0.8} + \sqrt{\frac{\pi}{2}}\sqrt{n}$ of (\ref{eqn:upperbound}).

\begin{figure}[t]
\centerline{\includegraphics[width=0.6\linewidth]{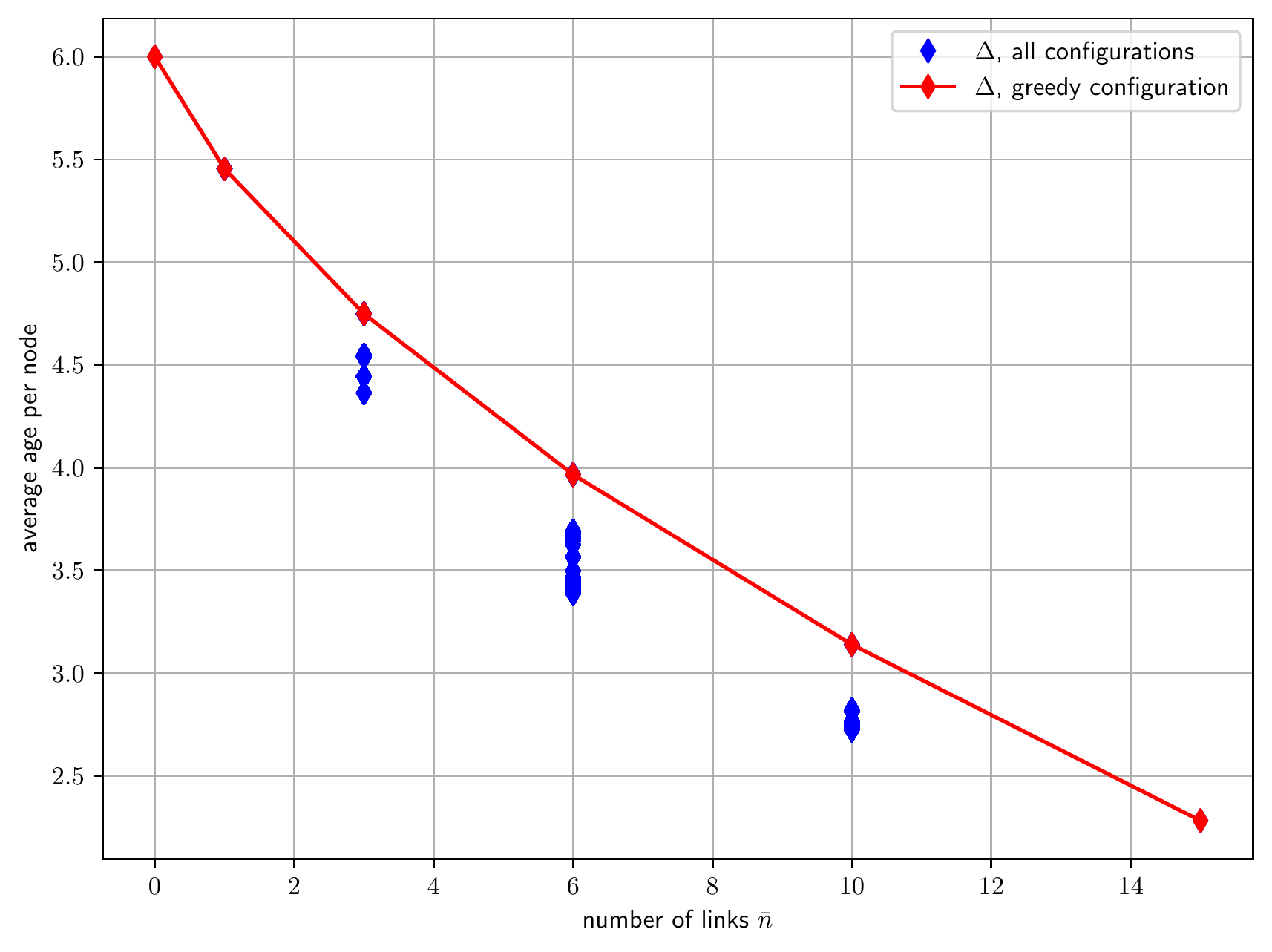}}
\caption{Blue dots represent the average age of a network of $n=6$ nodes for all $8480$ network configurations with $\Bar{n}=\binom{k}{2},k\in \{1,\ldots,n\}$ links. The red dots represent the average age when the links are added in accordance with the greedy approach.}
\label{fig:allconfig}
\end{figure}

Next, we consider the case of fully connected networks. In Fig.~\ref{fig:allconfig}, the blue points correspond to the average age of network of $n=6$ nodes for all $\sum_{k=1}^{n} \binom{\binom{n}{2}}{\binom{k}{2}}=8480$ network configurations that are possible with $\Bar{n}$ links, where we have picked $\Bar{n}=\binom{k}{2}$, $k\in\{1,\ldots,n\}$. The red line corresponds to the average age when the links are placed according to the greedy strategy, which seems to outperform all other link placements in this case. 

\begin{figure}[t]
\centerline{\includegraphics[width=0.55\linewidth]{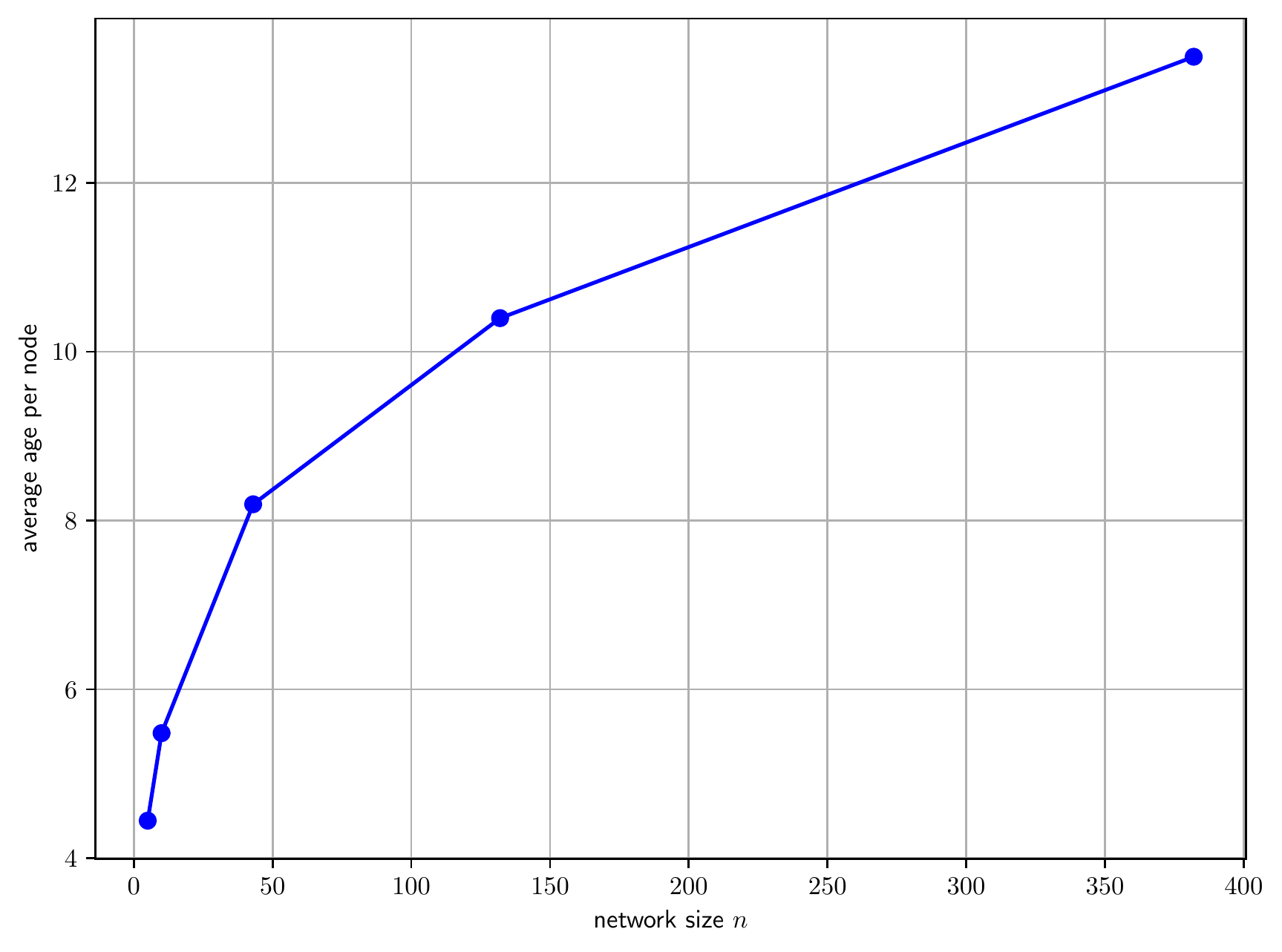}}
\caption{Average age in fully connected network with $n \log{n}$ jammers placed in accordance with the greedy method.}
\label{fig:nlogn_FC}
\vspace*{-0.4cm}
\end{figure}

Fig.~\ref{fig:nlogn_FC} shows the average plot as a function of the network size $n$ when the number of jammers is $\Tilde{n}=n\log{n}$, placed in accordance with the greedy strategy. The number of available links $\Bar{n}=\binom{n}{2}-\Tilde{n}$ will be of the form $\binom{k}{2}+c$, $c\in\{0,1,\ldots,k-1\}$, and we pick values of $n$ for which $c=0$ to have age increase consistently with $n$, since the graph increases consistently only for fixed values of $c$. We observe that the graph increases as $2 \log{n}$ with $k\approx \sqrt{n^2-2n\log{n}} \approx n-\log{n}$, as predicted in Lemma~\ref{lemma:FC_nlogn}. 

\begin{figure}[t]
\centerline{\includegraphics[width=0.55\linewidth]{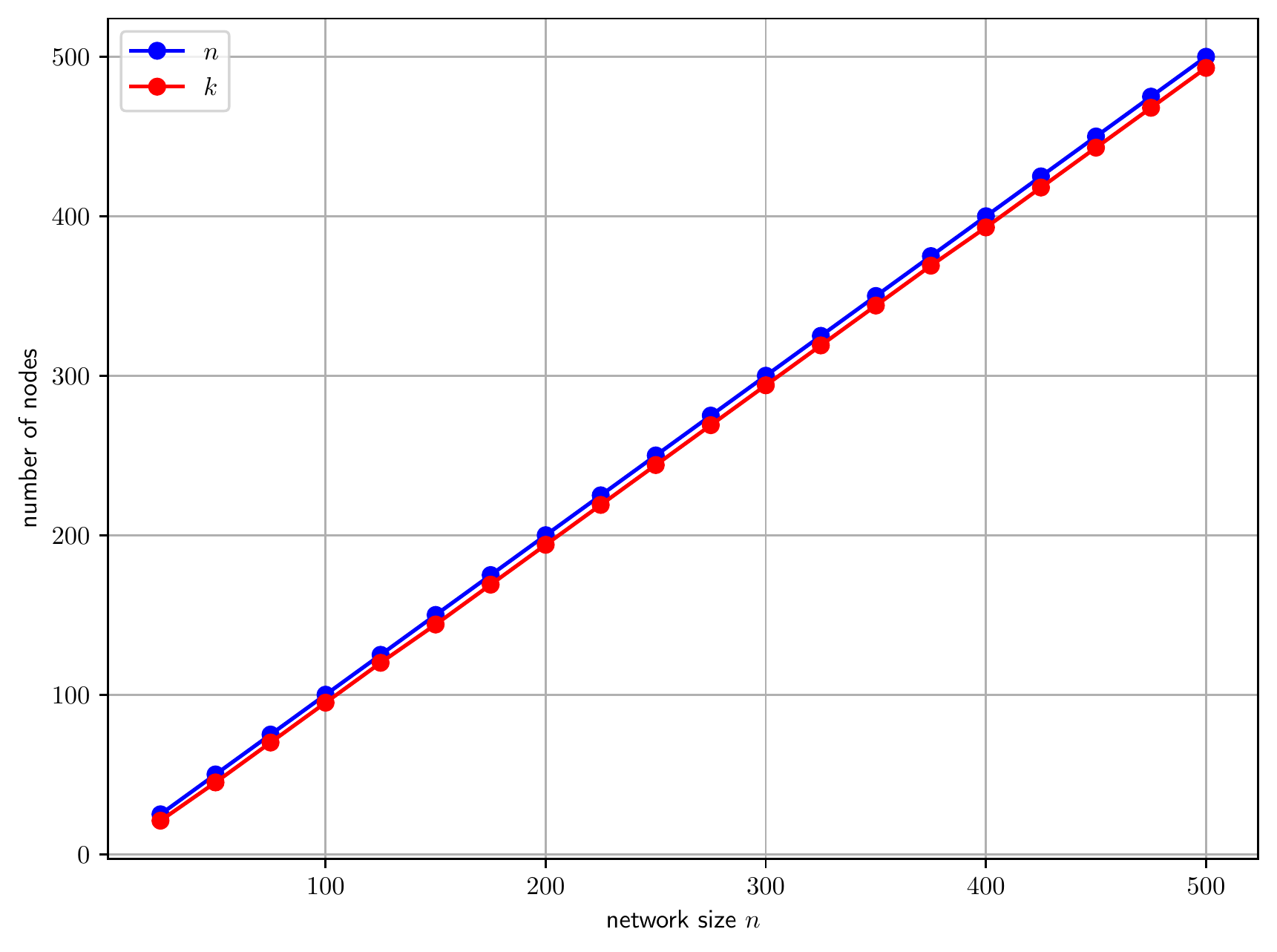}}
\caption{The blue line corresponds to the total number of nodes $n$ in the network, and the red line corresponds to the number of nodes $k$ that are left as part of the mini-FC when $n \log{n}$ jammers are placed on fully connected network in accordance with the greedy method.}
\label{fig:k_at_nlogn_FC}
\end{figure}

\begin{figure}[t]
\centerline{\includegraphics[width=0.55\linewidth]{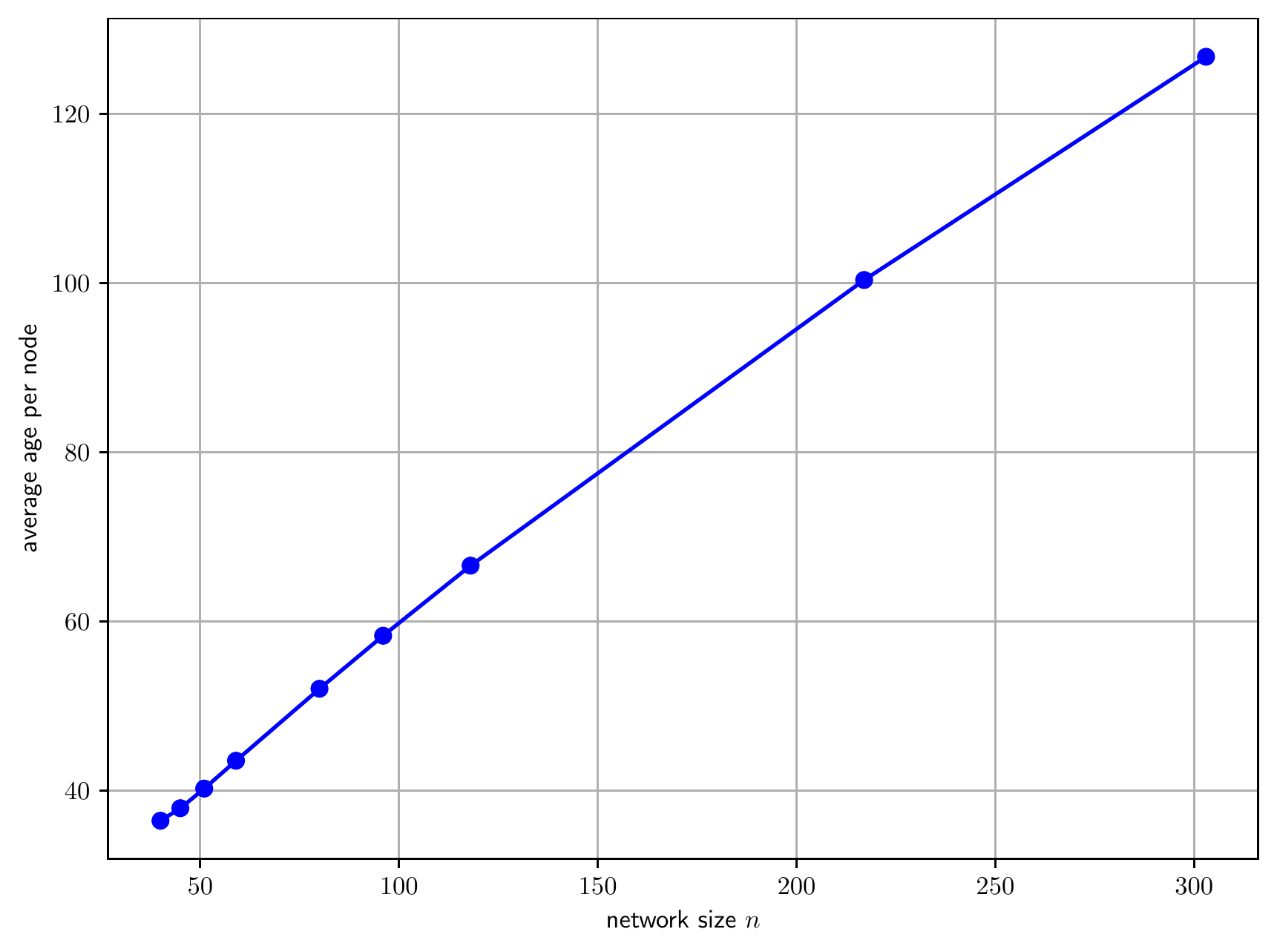}}
\caption{ Average age in fully connected network with $n^{1.8}$ jammers placed in accordance with the greedy method.}
\label{fig:FC_averageage_alpha8}
\vspace*{-0.4cm}
\end{figure}

Fig.~\ref{fig:k_at_nlogn_FC} shows the plot of $k$ and $n$ in red and blue line, respectively, where $k$ and $n$ are related as $\binom{k}{2}+c=\binom{n}{2}-n\log{n}$, $c\in\{0,1,\ldots,k-1\}$, i.e., $k$ is the largest size of the mini-FC with $n \log{n}$ jammers. We observe that the difference between the two lines is very small, since they only differ by a $\log{n}$ term. 

Finally, Fig.~\ref{fig:FC_averageage_alpha8} shows the average age plot as a function of the network size $n$ when the number of jammers is $\Tilde{n}=n^{1.8}$, choosing values of $n$ that give $c=0$ as before. The average age increases steeply owing to the presence of a larger number of jammers, and the graph looks almost linear because as $\alpha$ approaches $1$, $n^{\alpha}$ begins to appear linear. 

\section{Conclusion}

We first studied the effects of jamming attacks on the average age of a ring network. We showed that when the number of jammers $\tilde{n}$ scales as a fractional power of the network size $n$, i.e., $\tilde n= cn^\alpha$, the average version age scales as $\sqrt{n}$ when $\alpha \in \left[0,\frac{1}{2} \right)$, and as $n^{\alpha}$ when $\alpha \in \left[\frac{1}{2},1\right]$, implying that the version age with gossiping is robust against up to $\sqrt{n}$ jammers in a ring network, since version age of a ring network without any jammers scales as $\sqrt{n}$. Along the way, we studied average version age in line networks as well. We then studied the fully connected gossip network, where we derived a greedy approach to place $\Tilde{n}$ jammers with the goal to maximize the age of the resultant network, which involved using the $\Tilde{n}$ jammers to isolate the maximum possible nodes, thereby consolidating all links into a single mini-fully connected network. We showed in this network that the average version age scales as $O(\log{n})$ when $\Tilde{n}=O(n\log{n})$, and as $O(n^{\alpha-1})$, $1<\alpha\leq2$, when $\Tilde{n}=O(n^{\alpha})$, implying that the network is robust against $n\log{n}$ jammers, since version age of a fully connected network without jammers scales as $\log{n}$. Our work shows that connectivity improves resilience against jamming attacks and preserves timeliness of disseminated information; while the ring network (the lowest end of connectivity) is able to neutralize up to $\sqrt{n}$ jammers, the fully connected network (the highest end of connectivity) is able to neutralize up to $n \log n$ jammers, in an $n$-user gossip network. 

\bibliographystyle{unsrt}
\bibliography{IEEEabrv,ref_priyanka}
\end{document}